\documentclass{article}

 \usepackage[dblblindworkshop, final]{neurips_2025}
\workshoptitle{AI for Accelerated
Materials Discovery (AI4Mat) Workshop}

\usepackage[utf8]{inputenc} 
\usepackage[T1]{fontenc}    
\usepackage{hyperref}       
\usepackage{url}            
\usepackage{booktabs}       
\usepackage{amsfonts}       
\usepackage{nicefrac}       
\usepackage{microtype}      
\usepackage{xcolor}         

\usepackage[version=3]{mhchem}
\usepackage{subcaption}
\usepackage{graphicx}
\usepackage{float}
\usepackage{multirow}
\usepackage{array}
\usepackage{amsmath}
\usepackage{booktabs}
\usepackage{amsmath}
\usepackage{amssymb}
\usepackage{algorithm}
\usepackage{algorithmic}

\title{Migration as a Probe: A Generalizable Benchmark Framework for Specialist vs. Generalist Machine-Learned Force Fields}
\workshoptitle{AI for Accelerated Materials Discovery (AI4Mat) Workshop}

\author{%
  Yi Cao  \\ 
  Johns Hopkins University\\
  Baltimore, MD 21218 \\
  \texttt{ycao73@jh.edu} \\
  \And
  Paulette Clancy* \\
  Johns Hopkins University\\
  Baltimore, MD 21218 \\
  \texttt{pclancy3@jhu.edu} \\
}

\begin{document}

\maketitle

\begin{abstract}
Machine-learned force fields (MLFFs), particularly pre-trained foundation models, are revolutionizing computational materials science by enabling \textit{ab initio}-level accuracy at the length- and time-scales of classical molecular dynamics (MD). However, their rapid proliferation presents a critical strategic question: Should researchers train bespoke ``specialist'' models from scratch, fine-tune large ``generalist'' foundation models, or employ hybrid approaches? The trade-offs in data efficiency, predictive accuracy, computational cost, and susceptibility to out-of-distribution failure remain poorly understood, as does the fundamental question of how different training paradigms affect learned physical representations.

Here, we introduce a systematic benchmarking framework that addresses this question using defect migration pathways, evaluated via Nudged Elastic Band trajectories, as diagnostic probes that simultaneously test interpolation and extrapolation capabilities. Using Cr-doped \ce{Sb2Te3} as a technologically relevant 2D material case study, we benchmark multiple training strategies within the MACE architecture across equilibrium, kinetic (atomic migration), and mechanical (interlayer sliding) properties. 

Our key findings reveal that while all models adequately capture equilibrium structures, their predictions for non-equilibrium processes diverge dramatically. Targeted fine-tuning substantially outperforms both from-scratch and zero-shot approaches for kinetic properties, but induces catastrophic forgetting of long-range physics. Critically, analysis of learned representations shows that different training paradigms produce fundamentally distinct, non-overlapping latent space encodings, suggesting they capture different aspects of the underlying physics.

This work provides practical guidelines for MLFF development and establishes migration-based probes as an efficient, broadly applicable strategy for distinguishing model quality. We hope that this approach will offer a diagnostic framework that links performance to learned representations, paving the way for more intelligent, uncertainty-aware active learning strategies.
\end{abstract}
\section{Introduction}

The discovery and design of novel two-dimensional (2D) van der Waals (vdW) materials~\citep{novoselov_2d_2016, manzeli_2d_2017, guo_stacking_2021, liu_roadmap_2024} continues to drive innovation in spintronics~\citep{he_topological_2022}, quantum devices~\citep{qian_van_2024, song_electron_2018, nayak_microsoft_2025}, and energy technologies~\citep{jin_flexible_2019,gautam_creation_2024}. A particularly powerful strategy for tuning the properties of these layered materials involves doping—the insertion of guest atoms between vdW layers~\citep{novoselov_2d_2016, whittingham_lithium_2004}. This approach has enabled remarkable advances, from lithium-ion batteries~\citep{winter_insertion_1998, hu_lithium_2024} to the engineering of magnetic semiconductors~\citep{gibertini_magnetic_2019} and quantum anomalous Hall systems~\citep{tokura_magnetic_2019}.

Among 2D materials, topological insulators such as antimony telluride (\ce{Sb2Te3})~\citep{zhang_topological_2009, guo_roadmap_2015} represent an especially promising class for doping studies. The doping of transition metals has proven effective across multiple topological systems: iron in \ce{Bi2Se3} creates ferromagnetic order~\citep{checkelsky_trajectory_2014, kulbachinskii_thermoelectric_2012, haazen_ferromagnetism_2012}, manganese in \ce{Bi2Te3} enables tunable magnetic properties~\citep{klimovskikh_tunable_2020, fu_electronic_2021}, and vanadium in \ce{Sb2Te3} modifies electronic structure~\citep{sun_sb2te3_2023, zhang_quantum_2021}. The doping of chromium (Cr) into \ce{Sb2Te3}, the focus of this work, offers a compelling route to engineer its magnetic and topological properties for spintronic applications~\citep{cortie_creating_2020,han_effect_2013,de_a_deus_magnetic_2021}. Predicting the stability and dynamics of these complex systems is essential for guiding experimental synthesis, a task for which large-scale Molecular Dynamics (MD) simulations are indispensable~\citep{Zhang2023TuningStudy,thompson_lammps_2022}, provided a suitably representative force field is available. However, the atomic-scale processes governing functionality—including dopant migration and thermal stability—occur on time and length scales far beyond the reach of first-principles methods like Density Functional Theory (DFT)~\citep{kohn_self-consistent_1965, perdew_generalized_1996, sohier_density_2017,choudhary_high-throughput_2017}.

Machine-Learned Force Fields (MLFFs)~\citep{unke_machine_2021, ssmith_ani-1_2017, chen_accurate_2017, batatia_mace_nodate,vandermause_active_2022, batzner_e3-equivariant_2022, bartok_gaussian_2010,zeng_deepmd-kit_2023, wines_chips-ff_2025, musaelian_learning_2023, chen_universal_2022} have emerged as a powerful solution to this accuracy-versus-cost dilemma. The recent development of large-scale, pre-trained "foundation models" such as CHGNet~\citep{deng_chgnet_2023}, M3GNet/MEGNet~\citep{chen_graph_2019}, MACE-MP~\citep{batatia_mace_nodate}, MatterSim~\citep{yang_mattersim_2024}, and UMA~\citep{wood_uma_2025} marks a paradigm shift~\citep{merchant_scaling_2023}. This shift presents researchers with a critical and unresolved dilemma: for a specific system, is it more effective to invest significant resources to build a robust \textit{specialist} model from scratch~\citep{zhang_deep_2018, smith_approaching_2017, chmiela_sgdml_2017}, or to adapt a \textit{generalist} foundation model through fine-tuning? While generic benchmarks like Matbench~\citep{dunn_benchmarking_2020} validate their broad utility, their reliability for specific, high-fidelity applications is an open question~\citep{friederich_machine-learned_2021, deringer_machine_2019}. A valid concern is that these models may fail to capture nuanced interactions governing critical processes~\citep{schran_machine_2020, kapil_assessment_2022, grisafi_transferable_2018}. Fine-tuning is a promising solution, but naïve strategies can lead to \textit{catastrophic forgetting}, where general knowledge is erased. Key questions of data efficiency~\citep{janet_designing_2019, lookman_active_2019}, stability~\citep{zuo_performance_2020, jinnouchi_phase_2020}, and transferability~\citep{zeni_data_2021, vandermause_on--fly_2020} for non-equilibrium processes remain~\citep{bernstein_quantifying_2019, zhang_active_2019}.

Furthermore, when an MLFF-driven simulation fails, the underlying cause is often treated as a "black box." The default response—indiscriminately adding more data—is a brute-force approach that wastes resources. We propose that the key to overcoming this is to diagnose failure using physical probes. Specifically, we argue that migration pathways, evaluated via Nudged Elastic Band (NEB) trajectories~\citep{miskin2025low, ruttinger_protocol_2022}, provide a powerful and generalizable probe that simultaneously tests interpolation and extrapolation performance. Such migration-based probes impose stricter requirements than equilibrium-only tests, offering a sharper distinction between models and a path toward more intelligent, uncertainty-aware active learning.

This work addresses these gaps by presenting a systematic benchmark of bespoke training and fine-tuning strategies for the case study of Cr-doped \ce{Sb2Te3} using the MACE architecture. By combining latent-space analysis with migration-based probes, we move beyond traditional accuracy metrics to evaluate the dynamic stability and kinetic predictions of each model. Our study aims to serve as both a practical guide for selecting MLFF training strategies and a diagnostic framework for designing more data-efficient learning loops. To this end, we address the following key questions:
\begin{enumerate}
    \item How do bespoke MLFFs, trained from scratch, compare against a large pre-trained foundation model in terms of accuracy, stability, and data efficiency for simulating Cr-doped \ce{Sb2Te3}?
    \item What are the performance and stability trade-offs for fine-tuning strategies?
    \item Can migration pathways serve as a generalizable and efficient probe to benchmark specialist versus generalist models, and can latent-space analysis reveal signatures of failure to guide data acquisition?
\end{enumerate}

\section{Methodology}
We employed a systematic approach to benchmark Machine-Learned Force Fields (MLFFs) for Cr-doped \ce{Sb2Te3}
~\footnote{We chose this material system because it serves as an ideal testbed: its anisotropic 2D structure offers distinct migration environments—in-gap diffusion within quintuple layers (in-distribution) and interlayer migration across vdW gaps (out-of-distribution). This duality enables efficient sampling from equilibrium to high-energy transition states, addressing technologically relevant processes where both thermodynamic stability and kinetics are critical for doping engineering.}, comparing specialist models trained from scratch against fine-tuned foundation models.

\subsection{Data Generation and Model Training}
Reference data were generated using Density Functional Theory (DFT) calculations employing the PBE exchange-correlation functional~\citep{perdew1996generalized}. The dataset comprises approximately 20,000 atomic configurations of a $4\times 2 \times 1$ supercell of \ce{Sb2Te3} 
obtained from \textit{ab initio} molecular dynamics (AIMD) simulations conducted at three temperatures (300~K, 600~K, and 1200~K) using a 1~fs integration timestep.

All machine learning force field models were trained using the MACE architecture,~\citep{batatia_mace_nodate} which is an equivariant message-passing neural network designed for accurate representation of atomic interactions and many-body correlations.

Four distinct training strategies were evaluated:
\begin{enumerate}
\item \textbf{Scratch}: MACE model trained exclusively on our AIMD dataset
\item \textbf{Foundation}: Pre-trained MACE-MP model used without fine-tuning
\item \textbf{FT-600K}: Foundation model fine-tuned on 5\% of data from 600K trajectories
\item \textbf{FT-Multi\_T}: Foundation model fine-tuned on multi-temperature data
\end{enumerate}

\subsection{Evaluation Framework}
Models were assessed through: (i) MD simulations at 300K and 600K to evaluate structural stability and transport properties, (ii) nudged elastic band calculations for migration barriers, and (iii) representation learning analysis using t-SNE visualization of physically interpretable descriptors. Performance metrics included force/energy accuracy, dynamic stability, barrier prediction errors, and latent space clustering quality. Detailed computational parameters are provided in Appendix B.

\begin{figure}[htbp]
    \centering
    \includegraphics[width=0.7\textwidth]{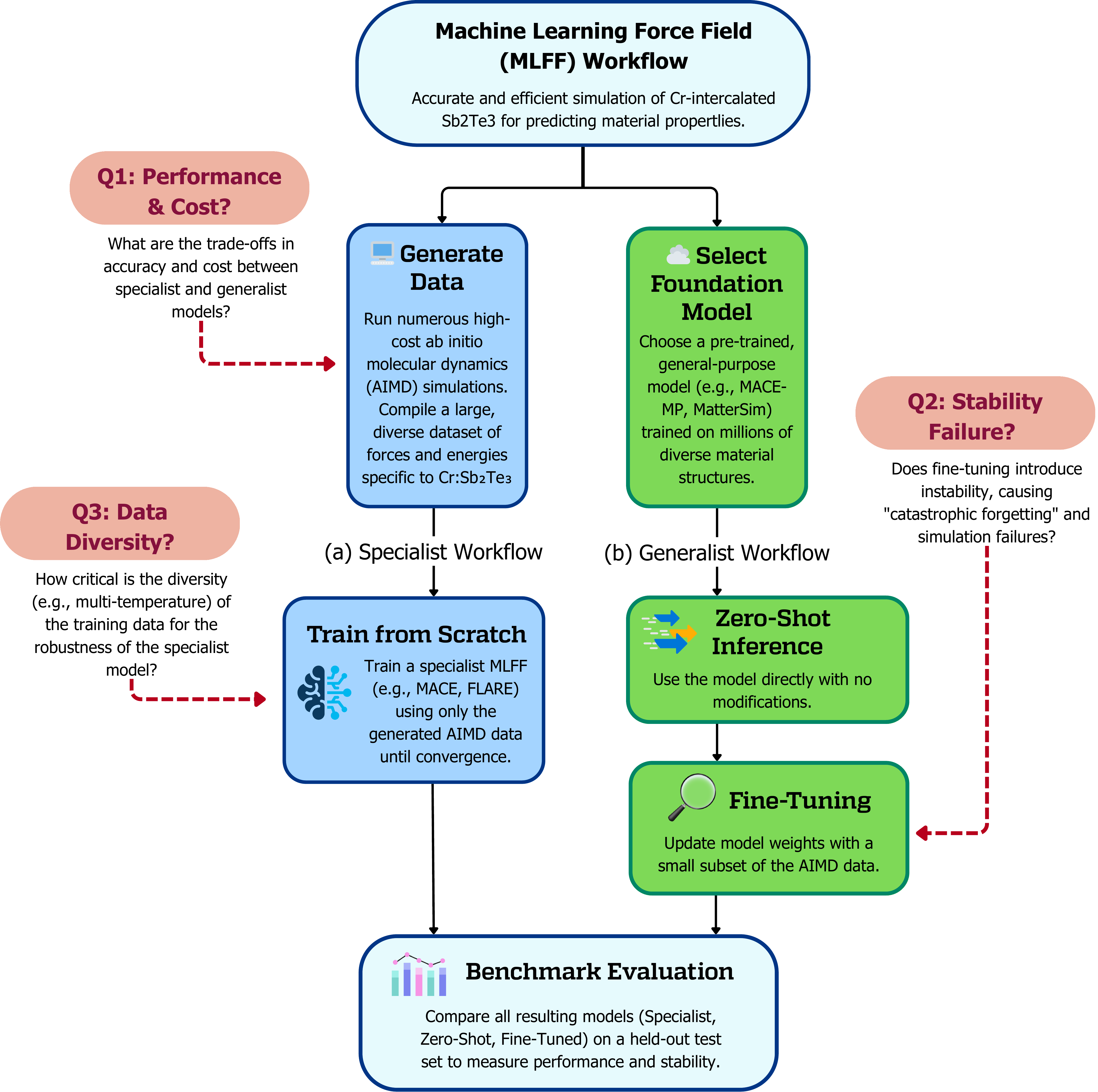}
    \caption{\textbf{Conceptual Diagram.} A flowchart illustrating two competing workflows: (a) training a specialist MLFF from scratch, and (b) fine-tuning a generalist foundation model. The diagram highlights the benchmarking questions this paper addresses.}
    \label{fig:conceptual_diagram}
\end{figure}

\section{Results and Discussions}

\subsection{Equilibrium Properties: Foundation for Comparison}
\begin{figure}[ht]
    \includegraphics[width=0.8\textwidth]{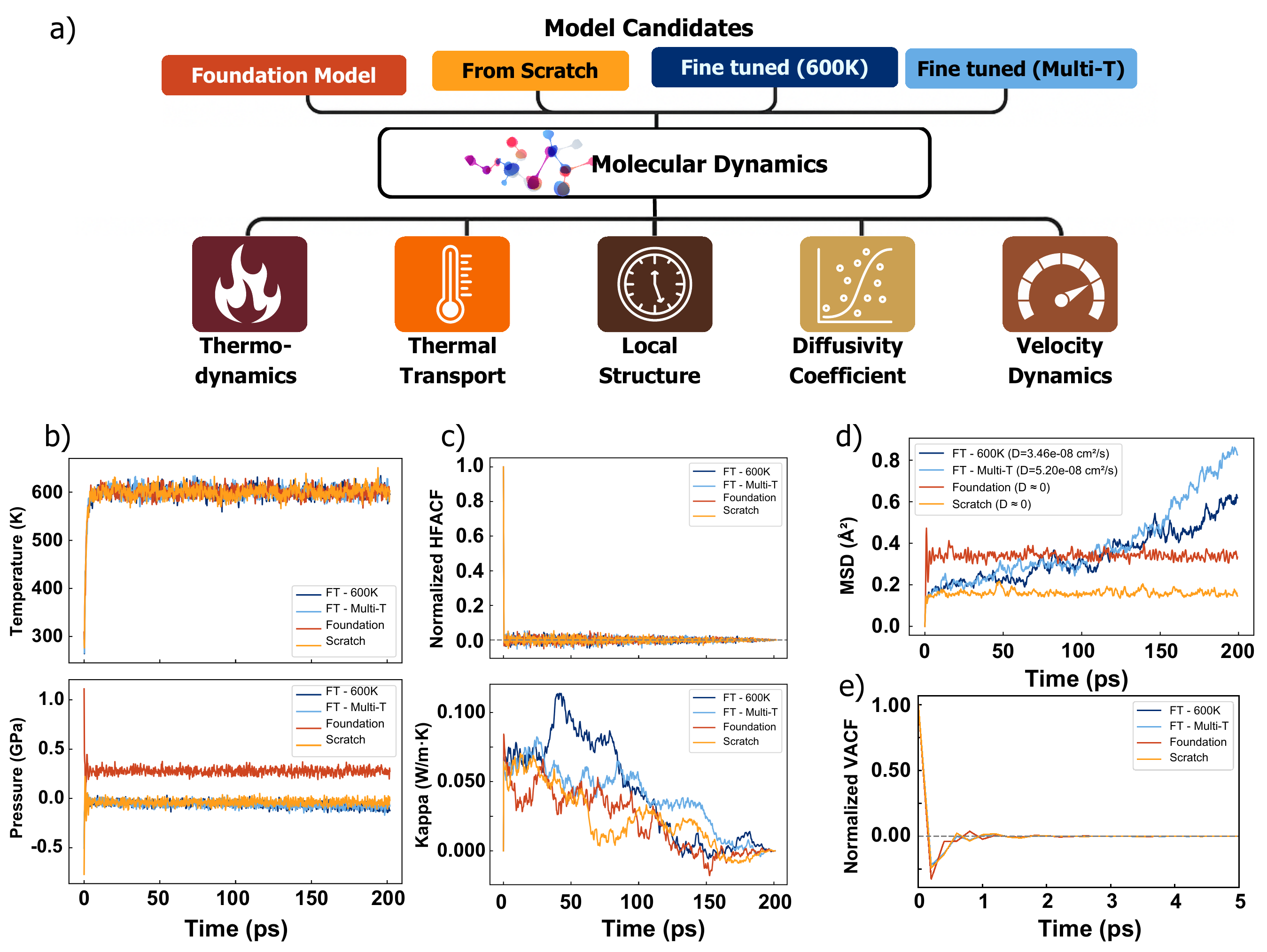}
    \centering
    \caption{\textbf{Comprehensive Benchmarking of MLFF Performance in Molecular Dynamics Simulations.} (a) A schematic of the unified evaluation protocol. Four candidate models—a zero-shot foundation model, a bespoke model trained from scratch, and two fine-tuned variants—are each used to drive a 200 ps MD simulation. The resulting trajectories are then subjected to a uniform set of post-processing analyses to evaluate key physical properties. (b) Thermodynamic stability, demonstrated by the evolution of temperature and pressure over the simulation, which remain stable around their target values for all models. (c) Thermal transport properties, showing the Heat Flux Autocorrelation Function (HFACF) and its running integral to compute thermal conductivity ($\kappa$).  (d) The Mean Squared Displacement (MSD), used to assess atomic mobility and calculate the diffusion coefficient. (e) The Velocity Autocorrelation Function (VACF), which describes the system's underlying dynamics.}
    \label{fig:ensemble}
\end{figure}

We considered four candidate models: a ``zero-shot'' foundation model, a bespoke model trained from scratch, and two fine-tuned variants of the foundation model. To make consistent evaluations to benchmark the performance of these models, we first assessed the performance of each training strategy via their performance in MD simulations (Fig.~\ref{fig:ensemble}a). 
This provides a strong test of each model's ability to generate not only thermodynamic and structural properties  but also accurately reproduce key dynamic, and transport properties. Each model was used to drive a 200~$ps$ simulation, with the resulting trajectories then subjected to the same set of post-processing analyses to evaluate their resulting key physical properties; shown in Figure~\ref{fig:ensemble}.

All MACE models---foundation, scratch, and fine-tuned variants---successfully reproduce the equilibrium structure of Cr-doped Sb$_2$Te$_3$, with RDF analysis showing excellent agreement with AIMD references for all atomic pair correlations (Fig.~\ref{fig:rdf}, see Appendix A for detailed analysis). While all models maintain stable thermodynamic ensembles at 600K (Fig.~\ref{fig:ensemble}c), the zero-shot foundation model exhibits a persistent pressure offset, likely due to its training on 0K equilibrium structures rather than finite-temperature configurations.

Despite similar performance on local structural (RDF) and short-time dynamic (VACF) properties (Fig.~\ref{fig:ensemble}e), the models diverge significantly in their predictions of long-timescale transport phenomena. Fine-tuned models predict higher diffusion coefficients than both foundation and scratch models (Fig.~\ref{fig:ensemble}d), with the multi-temperature fine-tuned variant showing the largest enhancement. This suggests that exposure to high-temperature training data creates a flatter potential energy  (PES) that persists even at lower temperatures.

Notably, thermal conductivity calculations reveal differences in how models capture collective vibrational modes. The foundation model's thermal conductivity decays rapidly toward zero, failing to sustain the long-range phonon correlations characteristic of the crystalline structure. In contrast, models trained or fine-tuned on system-specific data maintain these correlations, though the 600K-only model exhibits an anomalous peak suggesting potential structural instability. These results demonstrate that validating MLFFs solely on static structures and short-time dynamics is insufficient---a comprehensive assessment must include transport properties that probe long-range, collective phenomena.

\subsection{Non-Equilibrium Diffusion Pathways: The Critical Test of Generalization}

A rigorous test of the generalizability of a machine learning force field extends beyond its ability to reproduce thermodynamic properties. It must also accurately describe the potential energy surface (PES) in regions far from the training data's energetic minima, such as the high-energy transition states that govern kinetic processes. To this end, we evaluated the performance of our MACE model on the challenging task of predicting the diffusion barrier for a fundamental migration event in the Cr-doped Sb$_2$Te$_3$ system. This task is substantially more demanding than constant-temperature MD simulations (e.g., at 600~K), as it requires the model to capture not only accurate energies, but also the subtle curvature of the PES at a first-order saddle point---a stringent test of the model's learned representation of interatomic interactions.

\begin{figure}[ht]
    \includegraphics[width=0.8\textwidth]{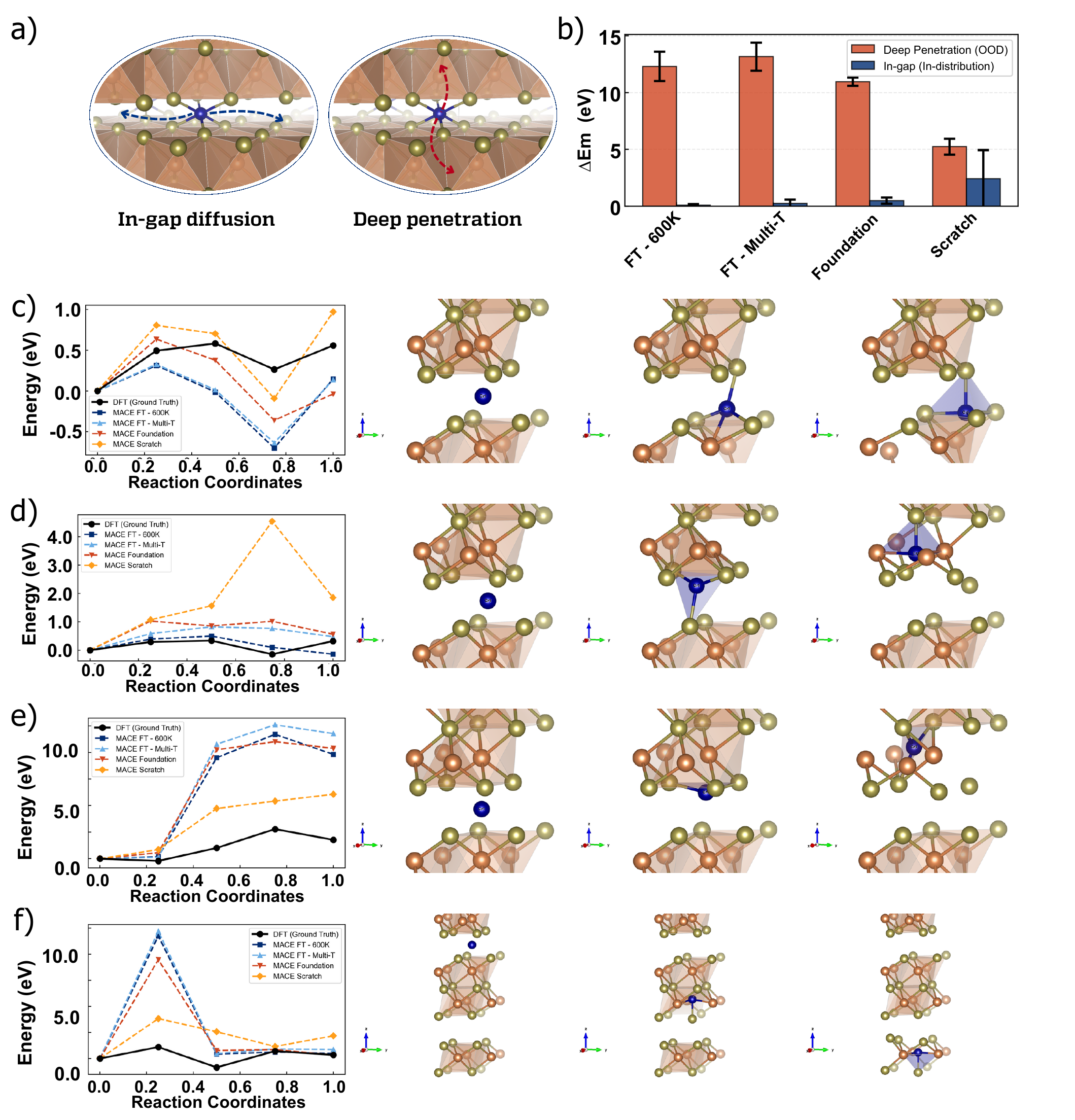}
    \centering
    \caption{\textbf{Comparative analysis of MLFF training strategies for predicting atomic migration barriers.} (a) Schematic illustration of the simulated Cr atom migration pathway between two stable sites within the Sb$_2$Te$_3$ bilayer. (b) Bar plot showing the migration energy prediction errors for various models, spanning bespoke training to advanced active learning approaches. (c--f) Comparison of the minimum energy pathway (MEP) profiles for the Cr migration process. The solid black line denotes the ground-truth DFT reference, while dashed lines represent predictions from MACE models trained with different strategies. 
    }
    \label{fig:barrier}
\end{figure}

\subsubsection{Local Migration Events}

To perform a direct and controlled evaluation of the learned PES for each model, we used the minimum energy pathway (MEP) obtained from our reference DFT nudged elastic band (NEB) calculation as a fixed geometric trajectory. By calculating the single-point energy for each of these pre-defined images, we can decouple the model's energetic accuracy from its structural relaxation behavior, providing a direct probe of its ability to describe the reaction pathway.

Our DFT calculations, serving as the ground-truth reference, establish a migration energy barrier ($E_{\mathrm{a}}$) of 0.34~eV for this process. The results, presented in Fig.~\ref{fig:barrier}d, reveal a clear hierarchy in performance and highlight the considerable impact of training strategy on the prediction of kinetic barriers. The two baseline strategies---training from scratch and using the zero-shot foundation model---demonstrate the fundamental challenges of specialization and generalization. The MACE Scratch model, despite being trained on a comprehensive in-house dataset, exhibits a catastrophic failure in predicting the barrier, overestimating the activation energy by over 4~eV. This is a classic signature of poor extrapolation. Even with tens of thousands of AIMD frames, the high-energy, low-probability configurations corresponding to the transition state are insufficiently sampled. Consequently, the model overfits to the more prevalent, near-equilibrium states and fails dramatically when asked to evaluate this critical, out-of-distribution rare event.

Conversely, the MACE Foundation model provides a qualitatively plausible prediction, capturing the smooth, convex shape of the energy barrier. This is a testament to the power of pre-training on millions of diverse structures; the model has learned a general understanding of physical interactions. However, it is quantitatively inaccurate, overestimating the barrier by approximately 0.7~eV---an unacceptably high amount---and misplacing the transition state along the reaction coordinate. This behavior is characteristic of a ``softened'' or ``averaged'' potential energy surface, a known trait of foundation models trained to generalize across vast chemical spaces. While broadly correct, the foundation model lacks the sharp, system-specific features of the true PES for Cr:Sb$_2$Te$_3$, effectively smoothing over the precise details required for high-fidelity kinetic predictions. The pronounced failures of these two baseline strategies underscore the necessity of a hybrid approach, motivating our investigation into fine-tuning.

\paragraph{The Critical Role of Task-Specific Fine-Tuning}

The dramatic improvement seen in the fine-tuned models underscores the necessity of specializing the foundation model's knowledge. The MACE FT--600K model, fine-tuned specifically on data generated at the target temperature of the migration event, achieves remarkable accuracy, with a barrier error of only 0.16~eV. This demonstrates that targeted fine-tuning effectively ``sharpens'' the softened PES of the foundation model. By providing a small but highly relevant set of in-domain data, we enable the model to learn the specific local atomic interactions required to accurately resolve the transition state structure and energy.

Intriguingly, the MACE FT--Multi-T model, which was fine-tuned on a larger and more diverse dataset spanning multiple temperatures, yields a less accurate barrier than the model trained at a specific temperature, here 600~K. While its predictions for the initial and final states (the thermodynamic endpoints) are accurate, its description of the kinetic barrier is a compromise, retaining some of the ``softened'' character of the original foundation model. This leads to an important insight: for MLFFs, more data is not axiomatically better. The relevance of the fine-tuning data to the specific task is paramount. For predicting thermodynamic properties, a multi-temperature dataset is superior. For predicting a specific kinetic process, a dataset rich in configurations relevant to that process's temperature regime is more effective. These findings collectively highlight a critical challenge in the practical application of MLFFs. The failure of the from-scratch model illustrates the difficulty of adequately sampling rare events, while the inaccuracies of the foundation model reveal the limitations of a purely generalist approach.

The success of targeted fine-tuning points the way forward, but the divergent results of the 600K and Multi-T models prove that the data selection process is non-trivial. This motivates the need for more advanced methods, such as the uncertainty-guided active learning explored in our work, which can intelligently identify and acquire the most informative data to improve model robustness and accuracy in a data-efficient manner.


To probe the generalization capabilities of the different MLFF training strategies, we evaluated their performance on two distinct Cr migration pathways (Fig.~\ref{fig:barrier}a) with fundamentally different characteristics. The first pathway, \emph{in-gap diffusion}, involves Cr migration within the vdW gap between Sb$_2$Te$_3$ quintuple layers, with low energy barriers and configurations similar to the training data, thereby testing interpolative accuracy.

In contrast, the second pathway, \emph{deep penetration}, requires the Cr atom to move vertically from the vdW gap and penetrate directly into the interior of a QL, ultimately reaching a deeply buried site within the layer. This migration path is associated with significantly higher energy barriers and accesses high-distortion configurations that are not well-represented in the training data---making it an out-of-distribution (OOD) scenario with potentially more metastable configurations along the pathway. Testing both pathways allows us to evaluate model accuracy on familiar configurations and extrapolative power in challenging regions of the configuration space.

\subparagraph{In-Gap Diffusion: A Test of Interpolative Accuracy}

For the in-gap diffusion pathway (Fig.~\ref{fig:barrier}c--d), where the Cr atom moves within the vdW gap or just shallowly penetrates the interface, the atomic environments along the minimum energy path (MEP) are reasonably similar to the near-equilibrium states sampled during the AIMD simulations. In this regime, the performance hierarchy is clear. The fine-tuned models, particularly MACE FT--600K, demonstrate excellent agreement with the DFT reference, accurately predicting both the thermodynamic endpoints and the kinetic barrier (Fig.~\ref{fig:barrier}d). This success highlights that, when the task lies within the domain of the training data, fine-tuning effectively ``sharpens'' the generalist foundation model's potential energy surface (PES) to capture system-specific details. In contrast, the MACE Scratch model, despite being trained on the same data, fails significantly, underscoring its inability to learn the subtle energy differences required to resolve the transition state from a limited dataset.

\subparagraph{Deep Penetration: A Test of Extrapolative Robustness}

The ``deep penetration'' pathway provides a much more stringent test of model robustness. This path involves significant lattice distortion as the Cr atom pushes its way through the covalently bonded quintuple layer (QL), creating atomic environments that are far from those seen in the training data (Fig.~\ref{fig:barrier}a).

While the fine-tuned and foundation models remain accurate for the stable initial and final states (an interpolative task), they fail catastrophically in predicting the energy of the transition state, overestimating the barrier by a large margin. This is a classic example of out-of-distribution failure, where the inductive biases learned from near-equilibrium data do not generalize to highly distorted, high-energy configurations. The models have learned to be a ``materials expert'' for stable structures but remain a ``naïve physicist'' for unseen, strained states.

The MACE Scratch model, which performed least well of the four models on the in-gap pathway, yields the lowest barrier error for this out-of-distribution task (Fig.~\ref{fig:barrier}e, f). This is not because the scratch model is ``better''; rather, its globally inaccurate and likely unphysical PES happens, by chance, to be less pathologically incorrect in this specific high-energy region than the foundation model's PES. The pre-trained model's failure suggests that its supposedly general knowledge contains strong implicit assumptions about near-equilibrium physics that break down dramatically when extrapolating.

\begin{figure}[ht]
    \includegraphics[width=0.6\textwidth]{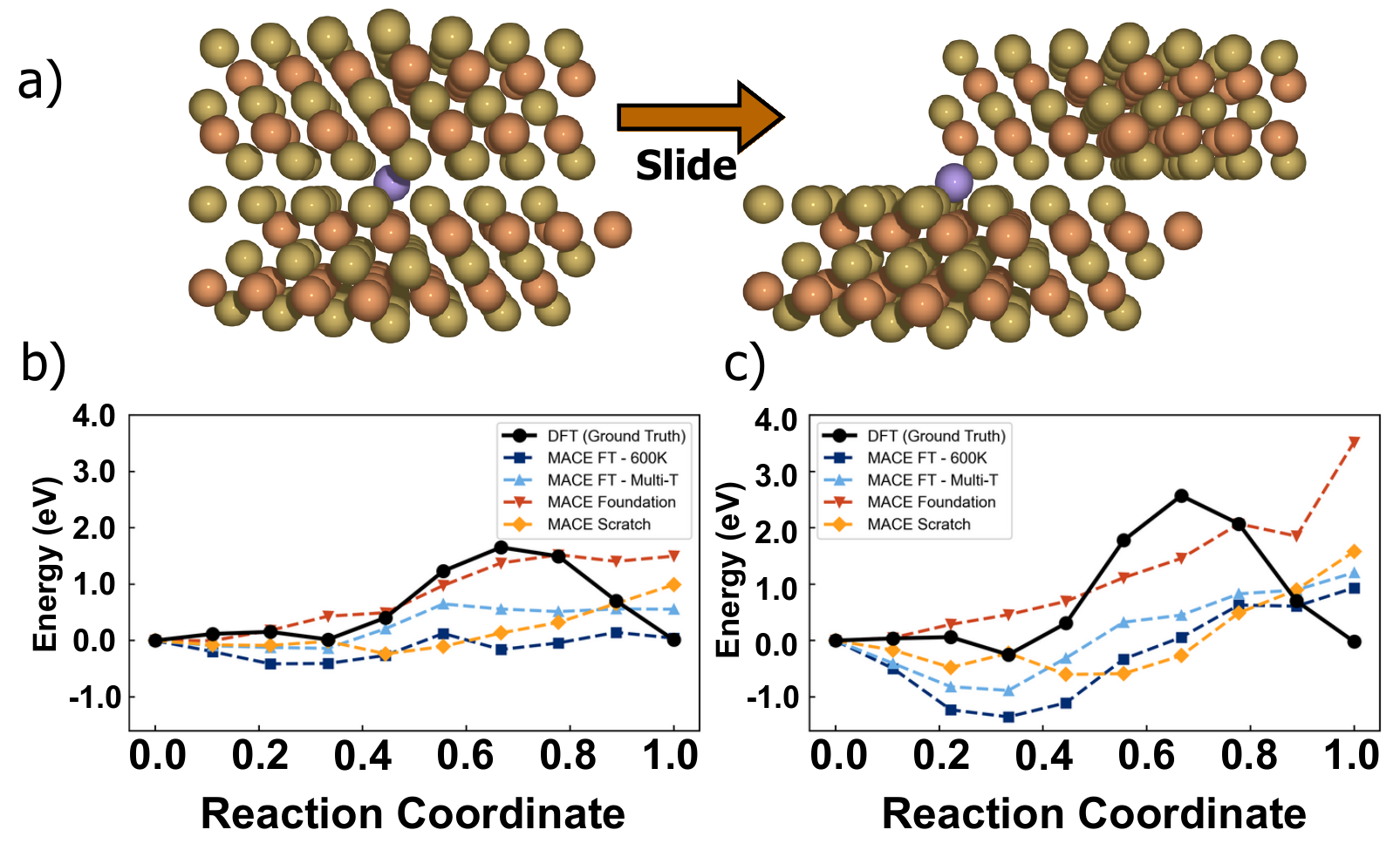}
    \centering
    \caption{\textbf{Evaluation of MACE models on collective lattice displacement: interlayer sliding in Sb$_2$Te$_3$.} (a) Schematic illustration of the bilayer sliding process in Sb$_2$Te$_3$, where the top layer (yellow/orange atoms) slides relative to the bottom layer along the crystallographic direction. The purple sphere indicates the position of a Cr dopant when present. (b) Energy barriers for interlayer sliding in pristine Sb$_2$Te$_3$ as calculated by DFT (black, ground truth) and various MACE models. The reaction coordinate represents the normalized sliding distance from the initial to final configuration. (c) Energy barriers for the same sliding process in Cr-doped Sb$_2$Te$_3$.}
    \label{fig:sliding_results}
\end{figure}

\subsection{Interlayer Sliding: Probing Robustness to Non-Local, Collective Displacements}

Having tested the models' response to local, high-distortion events, we evaluated their robustness to a different class of out-of-distribution challenge: large-scale, collective lattice displacements. To this end, we simulated the shearing of one Sb$_2$Te$_3$ layer relative to the other, a process governed by the subtle corrugations of the interlayer vdW potential energy surface (Fig.~\ref{fig:sliding_results}a). This task is particularly challenging for MLFFs as the local atomic environment of any given atom changes only minimally, while the global configuration undergoes a significant transformation that crosses periodic boundaries. The configurations along this sliding path were not present in the training data.

We first examined the case of pure, undoped Sb$_2$Te$_3$ (Fig.~\ref{fig:sliding_results}b). The results reveal a clear trade-off between the models. The MACE Foundation model provides the most reasonable estimate of the energy barrier's shape and magnitude, suggesting its vast pre-training on bulk crystals has endowed it with a better implicit understanding of such periodic, mechanical deformations. However, it fails to maintain translational symmetry, incorrectly predicting the final state to be higher in energy than the identical initial state. This energy drift is a clear artifact, indicating a failure to perfectly respect the periodic nature of the simulation cell under large displacements. The MACE Scratch model exhibits a similar version of this artifact.

Conversely, the fine-tuned models (MACE FT--600K and Multi-T) significantly underestimate the energy barrier. This suggests a compelling hypothesis: the process of fine-tuning, while ``sharpening'' the PES for local chemistry around the Cr dopant, has degraded the model's learned representation of the weaker, long-range interlayer physics inherited from the foundation model. The optimization has prioritized local accuracy at the expense of non-local, collective interactions.

This trend is largely mirrored in the Cr-doped system (Fig.~\ref{fig:sliding_results}c), confirming that this is a fundamental behavior of the models rather than an effect specific to the dopant. The failure of all models to perfectly capture both the barrier height and the endpoint periodicity highlights a key limitation of local-descriptor-based MLFFs. Phenomena like shearing, stacking faults, and dislocation glide are inherently non-local. While the models excel at describing local bonding and coordination, they can struggle to enforce long-range physical constraints that extend beyond their cutoff radius. This underscores the need for careful validation when using standard MLFFs to study mechanical properties and points towards future work in developing training sets that explicitly include these collective deformation modes or exploring architectures designed to capture long-range physics more effectively.

\subsection{Representation Learning Analysis}

\begin{figure}[ht]
    \includegraphics[width=\textwidth]{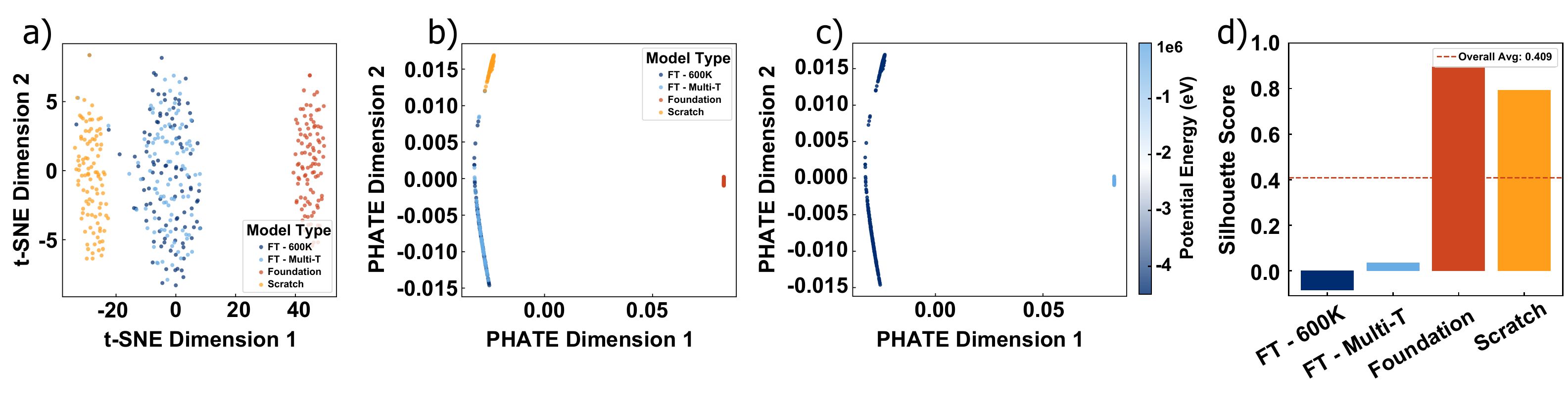}
    \centering
    \caption{\textbf{Representation Analysis Reveals How Fine-Tuning Aligns a Physical Manifold to Enable Generalization.} Projections of atomic environment descriptors from 600~K MD trajectories. 
    \textbf{(a)}~t-SNE projection shows that Foundation (red) and Scratch (orange) models produce highly separated representations, while fine-tuned models (blue) are intermediate. 
    \textbf{(b)}~PHATE projection reveals the system's continuous dynamical manifold. Note the clear separation between the brittle scratch model representation and the regularized fine-tuned models along the vertical axis (PHATE 2). 
    \textbf{(c)}~The same PHATE embedding colored by potential energy confirms that the manifold's geometry corresponds to the physical energy landscape. The foundation model's large energy offset indicates its nature as an uncalibrated prior.
    \textbf{(d)}~Average silhouette scores (from t-SNE) quantify the representational dissimilarity. The results demonstrate that fine-tuning succeeds by constraining a generalist representation to the specific physical manifold of the target system, a process that regularizes the model and enables accurate prediction of complex dynamics like diffusion.}
    \label{fig:tsne-phate}
\end{figure}

To elucidate the origins of the observed performance differences, we conducted a multi-faceted analysis of the learned representations. We projected the high-dimensional atomic environment descriptors from each model's 600~K MD trajectory into a two-dimensional space using two complementary techniques: t-distributed stochastic neighbor embedding (t-SNE)~\citep{maaten2008visualizing} to assess representational dissimilarity, and Potential of Heat-diffusion for Affinity-based Trajectory Embedding (PHATE)~\citep{moon2019visualizing} to reveal the underlying geometric structure of the system's dynamics (Fig.~\ref{fig:tsne-phate}).

The t-SNE projection (Fig.~\ref{fig:tsne-phate}a) confirms that models trained with different strategies learn qualitatively distinct encodings. The MACE Foundation and MACE Scratch models occupy well-separated regions of the latent space, a finding quantitatively supported by their high average silhouette scores (Fig.~\ref{fig:tsne-phate}d). This highlights the fundamental dissimilarity between the generalist prior and the specialist model trained from scratch. Crucially, the two fine-tuned models occupy an intermediate region, demonstrating that fine-tuning acts as a bridge between these two representational paradigms.

While t-SNE shows \textit{that} the representations are different, PHATE reveals \textit{why} it matters for physical prediction. The PHATE projection (Fig.~\ref{fig:tsne-phate}b) maps the continuous time-evolution of the system to a low-dimensional "dynamical manifold"—an arc-like structure whose geometry is intrinsically linked to the potential energy surface (Fig.~\ref{fig:tsne-phate}c). Along the primary axis of this manifold (PHATE 1), all specialist models (Scratch and FT) trace a similar region, indicating they have all learned the dominant, low-energy dynamics of the system.

The key insight comes from the separation along the second axis (PHATE 2). The scratch-trained model's representation is isolated from fine-tuned models in a distinct region (higher PHATE 2 values). This suggests it has learned a brittle, overfit representation, effectively memorizing a narrow path through the energy landscape without understanding the broader physical context. In stark contrast, the fine-tuned models are constrained to a different region of the manifold. This is the signature of regularization by pre-training; the foundation model's robust physical prior prevents the fine-tuned models from collapsing into the brittle state of the scratch model. Instead, they learn the system-specific dynamics \textit{within the context of a general and smooth physical manifold}.

This geometric difference in the latent space provides a direct mechanistic explanation for the models' performance on the diffusion task. Diffusion is a rare-event process that requires the model to generalize to out-of-distribution transition states. The scratch model's isolated manifold corresponds to a noisy and untrustworthy PES outside its training domain, leading to unphysical dynamics and a flattened MSD plot. Conversely, the fine-tuned models' robust, regularized manifold corresponds to a globally smoother and more reliable PES. This enables them to accurately predict the energy barriers and forces along the diffusion pathway, resulting in physically correct, linear MSD behavior. Fine-tuning succeeds not by creating a simple hybrid, but by aligning a general physical prior with the specific dynamical manifold of a target system, thereby enabling robust generalization to complex, long-timescale phenomena.

\section{Conclusion}

We benchmarked specialist (from-scratch) and generalist (foundation-based) machine-learned force fields (MLFFs) for Cr-doped Sb$_2$Te$_3$, emphasizing migration pathways as a diagnostic probe. NEB trajectories revealed that, while all models reproduce equilibrium structures, their performance diverges sharply for kinetic tasks. Migration thus provides a generalizable benchmark that tests both interpolation and extrapolation, exposing weaknesses invisible to equilibrium-only validation.  

Our results address the core questions posed in the introduction. First, both from-scratch and zero-shot foundation models fail for migration barriers: the former struggles with extrapolation, while the latter produces overly averaged potentials. Second, task-specific fine-tuning recovers kinetic accuracy, but at the cost of degraded performance for long-range physics such as interlayer sliding. Third, latent-space analysis shows these paradigms encode fundamentally distinct, non-overlapping physical representations, explaining hidden extrapolation failures.  

These findings reframe the specialist vs. generalist debate: the choice is not just about efficiency but about qualitatively different physics learned by each model. Foundation models excel at capturing broad chemical trends but require careful adaptation for system-specific kinetics, while specialist models offer more reliable local accuracy at the cost of transferability. Migration-based probes, coupled with latent-space diagnostics, offer a practical route to uncover hidden failures and guide uncertainty-aware active learning. 
\paragraph{Limitations and Future Work}
Our current study focuses on a single material system and migration mechanism as a proof-of-concept case study. Future work should extend this framework to multiple material classes, investigate uncertainty quantification methods for identifying unreliable predictions, and develop hybrid approaches that combine the transferability of foundation models with the precision of specialist training. Additionally, exploring active learning strategies guided by migration-based diagnostics could enable more efficient data collection for kinetic properties. This framework has the potential to generalize beyond the present case, enabling more robust and uncertainty-aware MLFF development for accelerated materials discovery.
\newpage
\appendix

\section{Detailed Computational Methods}
\subsection{First-Principles Reference Calculations}
All reference data were generated using Density Functional Theory (DFT) as implemented in the Quantum Espresso simulation package. We employed ultrasoft pseudopotentials for all elements (Cr, Sb, Te) with a plane-wave kinetic energy cutoff of 400 eV. The Brillouin zone was sampled using a Monkhorst-Pack k-point grid of $4 \times 4 \times 1$ for structural relaxations and Nudged Elastic Band (NEB) calculations. The Perdew-Burke-Ernzerhof (PBE) exchange-correlation functional was used, which we have found in prior investigations to provide a reliable balance of computational efficiency and accuracy for this class of materials.

The training dataset was generated from AIMD simulations performed in the NVT ensemble using a Langevin thermostat. These simulations covered a range of temperatures (300 K, 600 K, 1200 K) and Cr doping concentrations to ensure the model was exposed to a diverse set of thermodynamic and structural configurations. Each AIMD trajectory was run for 10 ps on a system of 120 atoms. To investigate the nature of atomic migration pathways, minimum energy paths (MEPs) and energy barriers were calculated using the Nudged Elastic Band (NEB) method. All initial, final, and intermediate configurations were considered to be converged when the forces on all unconstrained atoms fell below 0.01 eV/\AA.

\subsection{MACE Model Training Protocols}
\subsubsection{Training Hyperparameters}
All training and fine-tuning procedures were executed with a consistent set of hyperparameters to ensure fair comparison:
\begin{itemize}
    \item Optimizer: Adam
    \item Initial learning rate: $1\times10^{-3}$
    \item Batch size: 4
    \item Early stopping: Implemented based on validation set Force MAE
    \item Maximum epochs: 1000
    \item Validation split: 10\% of training data
\end{itemize}

\subsubsection{Fine-Tuning Details}
For the FT-600K model, we selected a representative 5\% subset (approximately 1,000 configurations) from the 600K AIMD trajectories. The subset was chosen to capture the full range of structural variations observed at this temperature, including both equilibrium fluctuations and transitional configurations.

For the FT-Multi\_T model, the training subset was composed equally from 300K, 600K, and 1200K trajectories, ensuring exposure to diverse thermal conditions. The multi-temperature dataset was designed to test whether broader thermodynamic sampling improves generalization.

The choice of 600K for single-temperature fine-tuning represents typical thermoelectric operating temperatures, reflecting realistic usage conditions. At roughly two-thirds of Sb$_2$Te$_3$'s melting point, this temperature captures significant thermal dynamics without structural deterioration.

Fine-tuning foundation models for specific chemical systems presents a fundamental challenge: how to adapt to new domains while preserving learned representations. This document discusses our fine-tuning strategy.

\subsubsection{Fine-tuning Underlying Mechanism}

In fine-tuning, the pre-trained model parameters $\theta_0$ are directly optimized on the target dataset $\mathcal{D}_{\text{target}}$:

\begin{equation}
\theta^* = \arg\min_\theta \mathcal{L}(\theta; \mathcal{D}_{\text{target}})
\end{equation}

\paragraph{Characteristics}
\begin{itemize}
    \item \textbf{Simple implementation}: Direct optimization on new data
    \item \textbf{Fast convergence}: High learning rate ($\alpha \sim 10^{-2}$)
    \item \textbf{Catastrophic forgetting}: Loss of original capabilities
    \item \textbf{Single objective}: Optimizes only for target domain performance
\end{itemize}

\subsubsection{Implementation Details}

\subsubsection{Fine-tuning Algorithm}
\begin{algorithm}
\caption{Fine-tuning}
\begin{algorithmic}
\STATE Initialize: $\theta \leftarrow \theta_{\text{foundation}}$
\FOR{epoch = 1 to $N_{\text{epochs}}$}
    \FOR{batch $\in \mathcal{D}_{\text{target}}$}
        \STATE $\mathcal{L} \leftarrow \text{MSE}(f_\theta(\text{batch}), y_{\text{batch}})$
        \STATE $\theta \leftarrow \theta - \alpha \nabla_\theta \mathcal{L}$
    \ENDFOR
\ENDFOR
\end{algorithmic}
\end{algorithm}

\subsubsection{Practical Considerations}

\paragraph{When to Use Fine-tuning}
\begin{itemize}
    \item Limited computational resources
    \item Domain-specific applications only
    \item Rapid prototyping requirements
    \item No need for generalization beyond the target dataset
\end{itemize}

\paragraph{Take-away}

Fine-tuning offers rapid adaptation to new datasets with relatively low computational cost. However, this comes at the expense of catastrophic forgetting, where the model loses its original generalization capabilities.

\section{Supplementary Materials}
\subsection{MACE Model Performance Evaluation by RMSE}

\begin{figure}[ht]
    \includegraphics[width=0.8\textwidth]{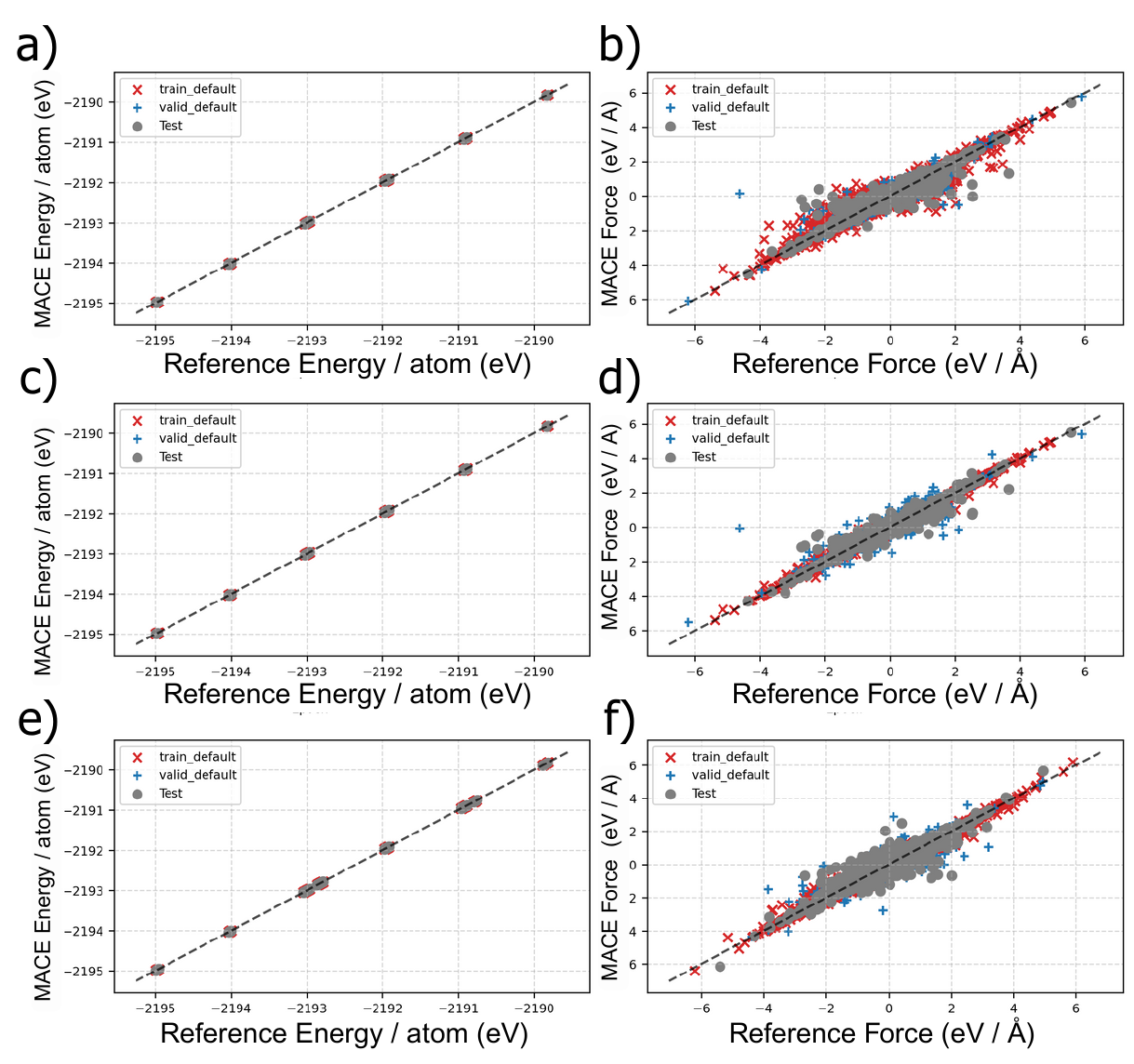}
    \centering
    \caption{\textbf{Parity plots comparing MACE-predicted energies and forces against DFT reference calculations for the test set.} (a, b) Model trained from scratch, (c, d) fine-tuned 600 K model (``FT-600K''), and (e, f) fine-tuned multi-temperature model (``FT-MultiT''). Parity plots for energies (a, c, e) demonstrate excellent agreement for all models, while force parity plots (b, d, f) show strong overall correlation but subtle differences in error distributions that are not distinguishable visually. These results highlight the necessity of quantitative and property-based benchmarks for robust model evaluation.}
    \label{fig:mace_parity}
\end{figure}

All developed Machine-Learned Force Fields (MLFFs) demonstrate high fidelity in predicting energies and forces, as illustrated by the parity plots in Figure~\ref{fig:mace_parity}, which compare model predictions to the reference DFT calculations. The performance of each model across the training, validation, and test sets is quantitatively summarized in Table~\ref{tab:mace_performance_detailed}. As detailed in the table, the models fine-tuned from the foundation model show a marked improvement over the model trained from scratch, with energy and force Root Mean Square Errors (RMSEs) being significantly reduced on the independent test set.

\begin{table}[htbp]
    \centering
    \caption{Performance metrics for MACE models across training, validation, and test sets. The FT-600K metrics are the mean $\pm$ standard deviation across three independent training runs with different random seeds.}
    \label{tab:mace_performance_detailed}
    \begin{tabular}{lcccc}
        \toprule
        \textbf{Model} & \textbf{Train RMSE F} & \textbf{Valid RMSE F} & \textbf{Test RMSE F} & \textbf{Test RMSE E} \\
         & (meV/\AA) & (meV/\AA) & (meV/\AA) & (meV/atom) \\
        \midrule
        From Scratch & 67.1 & 76.1 & 75.2 & 1.0 \\
        FT-600K & 20.7 $\pm$ 11.8 & 44.6 $\pm$ 2.9 & 37.2 $\pm$ 0.6 & 0.5 $\pm$ 0.0 \\
        FT-MultiT & 20.3 & 49.1 & 45.5 & 0.5 \\
        \bottomrule
    \end{tabular}
\end{table}

A deeper analysis of the training dynamics reveals important characteristics of the models. The model trained from scratch exhibits a relatively small gap between training (67.1 meV/\AA) and validation (76.1 meV/\AA) force RMSEs. While this suggests only moderate overfitting, its high error across all datasets indicates that it underfits the true physical interactions.

In contrast, the fine-tuned models display a more pronounced overfitting behavior, characterized by a large gap between their low training RMSEs and higher validation RMSEs. For instance, the FT-MultiT model shows a 142\% increase in force RMSE from the training (20.3 meV/\AA) to the validation set (49.1 meV/\AA). This aggressive fitting is expected when fine-tuning on smaller, more specialized datasets. The high standard deviation in the FT-600K model's training RMSE (11.8 meV/\AA) also suggests a sensitivity to initialization during the training process.

Despite the strong overfitting signals, the fine-tuned models generalize well to the test set, outperforming the scratch model by a significant margin. This analysis reinforces the central argument of our work: standard accuracy metrics like RMSE are insufficient for a comprehensive evaluation of MLFFs. While these metrics and the parity plots in Figure~\ref{fig:mace_parity} confirm the models' general accuracy, they do not capture critical performance on physical properties such as energy barriers and transport phenomena, necessitating the more extensive, property-driven benchmarks presented in the main text.

\subsection{Detailed Structural and Thermodynamic Analysis}
\begin{figure}[ht]
    \includegraphics[width=0.8\textwidth]{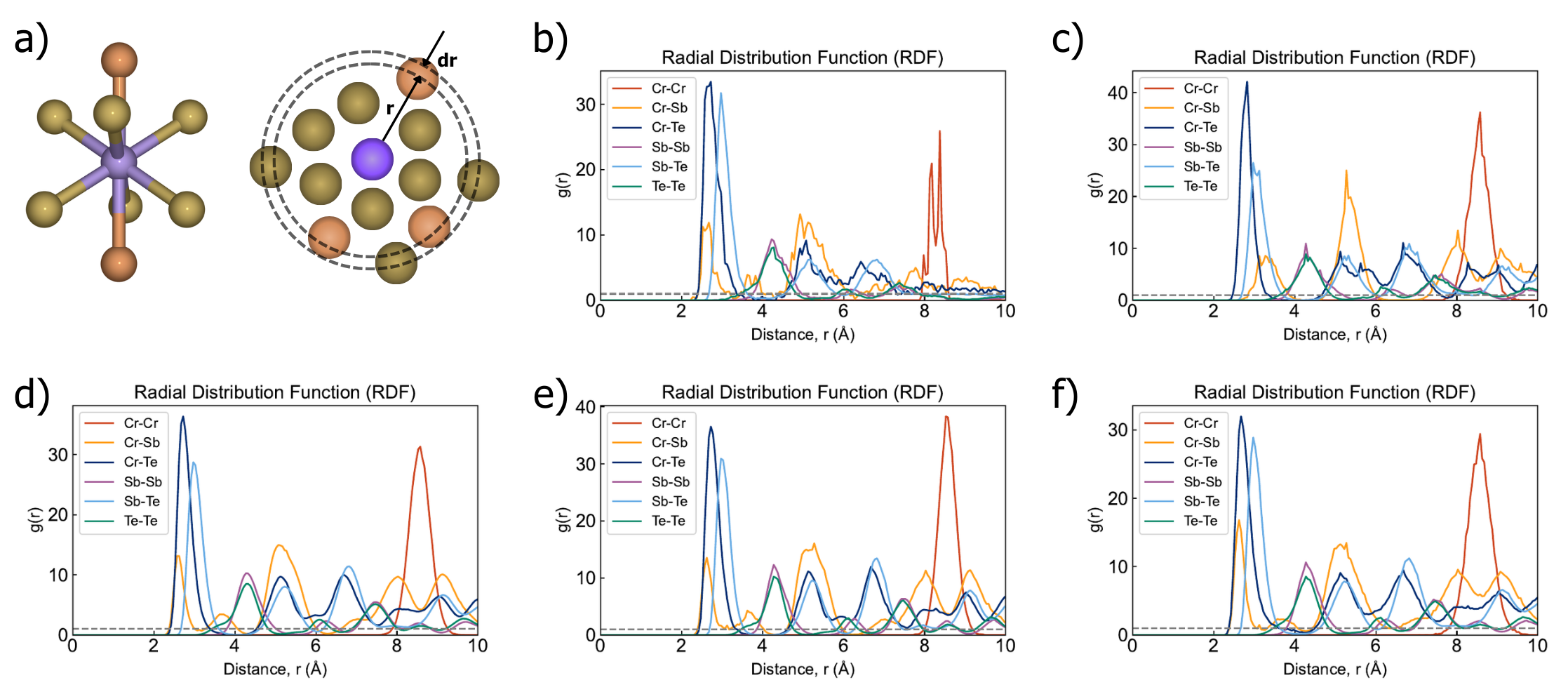}
    \centering
    \caption{\textbf{Radial distribution function (RDF) analysis of Cr-doped Sb$_2$Te$_3$ at 600 K.} (a) Schematic illustration of the local atomic environment around a Cr dopant (purple) in the Sb$_2$Te$_3$ lattice, showing the first coordination shell of Te atoms (gold) and second-shell Sb atoms (orange). The dashed circles indicate the radial distances used for RDF calculation. (b-f) Computed RDFs for different atom pairs from MD simulations: (b) Ab initio molecular dynamics (AIMD) ground truth reference, (c) MACE foundation model without fine-tuning, (d) MACE model trained from scratch on Cr-Sb$_2$Te$_3$ data, (e) MACE model fine-tuned with 600 K AIMD data (``FT - 600K''), and (f) MACE model fine-tuned with multi-temperature AIMD data (``FT - Multi-T''). All simulations were performed at 600 K with 5$\times$5$\times$1 supercells for 200 ps. The RDF peaks correspond to characteristic interatomic distances in the doped structure, with the first peak representing nearest-neighbor correlations.}
    \label{fig:rdf}
\end{figure}

The RDF analysis reveals that all MACE models, regardless of training strategy, successfully reproduce the key structural features of the Cr-doped Sb$_2$Te$_3$ system when compared to the AIMD ground truth (Fig.~\ref{fig:rdf}b).

All models correctly capture the primary coordination shells, with the first peak positions for Cr--Cr, Cr--Sb, and Cr--Te pairs occurring at approximately 3.0~\AA, 3.2~\AA, and 2.8~\AA, respectively, in excellent agreement with the AIMD reference. The peak heights and positions remain consistent across all training strategies (Figures~\ref{fig:rdf}c--f), indicating that the local atomic structure is well preserved regardless of whether the model was trained from scratch, used as a foundation model, or fine-tuned with temperature-specific data.

The primary observable difference between models lies in the smoothness of the RDF curves rather than their fundamental features. The scratch-trained model (Fig.~\ref{fig:rdf}d) exhibits slightly smoother RDF profiles compared to the fine-tuned variants (Figures~\ref{fig:rdf}e--f). This difference arises from practical computational considerations rather than fundamental accuracy: the scratch model, being more compact with fewer parameters, allowed for longer MD trajectories within the same computational budget, resulting in better statistical sampling. In contrast, the fine-tuned models, while maintaining larger parameter counts from their foundation architectures, required more computational resources per MD step, limiting the total simulation time and resulting in slightly noisier RDF profiles.

Notably, both fine-tuning strategies---whether using single-temperature (600~K) or multi-temperature AIMD data---produce nearly identical RDF profiles, suggesting that the structural representation of the material is robust to the temperature range of the training data. This indicates that, for structural properties, the choice of training strategy has minimal impact, with all approaches converging to similar descriptions of the local atomic environment. The preservation of accurate structural features across all models provides confidence that the learned interatomic potentials correctly capture the fundamental bonding characteristics of the Cr-doped Sb$_2$Te$_3$ system.

\subsection{Thermodynamic Ensemble Stability}
A detailed analysis of the pressure profiles reveals subtle but important differences between models. The zero-shot MACE Foundation model equilibrates to a slightly different average pressure than the models trained or fine-tuned on our in-house AIMD data. This is likely a consequence of the foundation model being trained on a vast dataset of materials at their 0K equilibrium volumes. A minor mismatch between the foundation model's predicted equilibrium lattice parameters and the true DFT values for our specific Cr:Sb$_2$Te$_3$ system at 600K manifests as a persistent non-zero average pressure in an NVT simulation. The fine-tuned and scratch models, having been trained explicitly on data from this system's true ensemble, do not exhibit this deviation.

\subsection{Transport Property Analysis}
The divergence in transport properties provides deeper insights into model behavior. The enhanced diffusivity in multi-temperature fine-tuned models likely stems from their exposure to high-temperature configurations approaching disordered or liquid-like states during training. By learning from these states, the model may have developed a potential energy surface that is slightly "flatter" or has lower barriers to diffusion, an effect that persists even at the 600K simulation temperature.

The thermal conductivity analysis reveals even more fundamental differences. The rapid decay of thermal conductivity in the foundation model indicates a failure to sustain long-range heat-carrying vibrational modes (phonons) specific to this crystalline structure. The anomalous peak observed in the 600K-only fine-tuned model's HFACF during the first 50 ps suggests potential structural instability or abrupt structural changes that alter phonon behavior. This highlights how different training strategies can lead to qualitatively different representations of collective phenomena, even when local properties appear identical.

\subsection{Molecular Dynamics Simulations}

\subsubsection{Simulation Protocol}

MD simulations were performed using the Atomic Simulation Environment (ASE) with the following protocol:

\begin{itemize}
    \item \textbf{Integrator}: Langevin dynamics for NVT ensemble or Nosé-Hoover for NPT ensemble
    \item \textbf{Timestep}: 1.0 fs
    \item \textbf{Friction coefficient}: $\gamma = 0.01$ fs$^{-1}$ (Langevin)
    \item \textbf{Barostat parameters}: $\tau_p = 1000$ fs, $P = 1$ bar (NPT only)
    \item \textbf{Equilibration}: 50,000 steps (50 ps)
    \item \textbf{Production}: 200,000 steps (200 ps) for property calculations
    \item \textbf{Sampling interval}: Every 100 steps for analysis
    \item \textbf{System size}: 5$\times$5$\times$1 supercell (2050 atoms)
\end{itemize}

Simulations were conducted at two temperatures: 300 K and 600 K, to assess model performance under different thermal conditions. Initial velocities were assigned according to the Maxwell-Boltzmann distribution with removal of center-of-mass motion and angular momentum.

\subsection{Property Calculations}

\subsubsection{Structural Properties}

The radial distribution function (RDF) $g(r)$ was computed for all unique atom pairs:

\begin{equation}
g_{\alpha\beta}(r) = \frac{V}{4\pi r^2 \Delta r N_\alpha N_\beta} \left\langle \sum_{i \in \alpha} \sum_{j \in \beta} \delta(r - r_{ij}) \right\rangle
\end{equation}

where $\alpha$ and $\beta$ denote atomic species, $V$ is the system volume, $N_\alpha$ is the number of atoms of species $\alpha$, and the angle brackets denote time averaging.

\subsubsection{Dynamic Properties}

The mean square displacement (MSD) was calculated for each atomic species:

\begin{equation}
\text{MSD}_\alpha(t) = \left\langle \frac{1}{N_\alpha} \sum_{i \in \alpha} |\mathbf{r}_i(t) - \mathbf{r}_i(0)|^2 \right\rangle
\end{equation}

Diffusion coefficients were extracted from the linear regime of MSD using the Einstein relation:

\begin{equation}
D_\alpha = \lim_{t \to \infty} \frac{\text{MSD}_\alpha(t)}{6t}
\end{equation}

The velocity autocorrelation function (VACF) was computed as:

\begin{equation}
C_v(t) = \frac{\langle \mathbf{v}(t) \cdot \mathbf{v}(0) \rangle}{\langle \mathbf{v}(0) \cdot \mathbf{v}(0) \rangle}
\end{equation}

\subsubsection{Thermodynamic Properties}

Average potential energy per atom, temperature, and volume (for NPT simulations) were calculated over the production phase:

\begin{equation}
\langle E \rangle = \frac{1}{N_{\text{frames}}} \sum_{i=1}^{N_{\text{frames}}} \frac{E_{\text{pot}}^{(i)}}{N_{\text{atoms}}}
\end{equation}

with corresponding standard deviations to assess thermal fluctuations.

\subsubsection{Transport Properties}

For thermal conductivity calculations, the heat flux vector was computed:

\begin{equation}
\mathbf{J} = \frac{1}{V} \left[ \sum_i e_i \mathbf{v}_i + \frac{1}{2} \sum_{i \neq j} (\mathbf{F}_{ij} \cdot \mathbf{v}_j) \mathbf{r}_{ij} \right]
\end{equation}

where $e_i$ is the per-atom energy, $\mathbf{v}_i$ is the velocity, $\mathbf{F}_{ij}$ is the force between atoms $i$ and $j$, and $\mathbf{r}_{ij}$ is their separation vector. The heat flux autocorrelation function (HFACF) was then computed for subsequent Green-Kubo analysis.

\subsection{Representation Learning Feature Extraction}
The following physically interpretable descriptors were extracted from MD trajectories:
\begin{enumerate}
\item \textbf{Energy landscape}: Total potential energy per atom
\item \textbf{Force fields}: 3N-dimensional force vectors for all atoms
\item \textbf{Structural order parameters}: Radial distribution histograms computed with 100 bins up to 5.0 \AA{} cutoff
\item \textbf{SOAP descriptors}: Smooth Overlap of Atomic Positions with:
  \begin{itemize}
  \item Cutoff radius: 5.0 \AA
  \item Number of radial basis functions: 8
  \item Maximum angular momentum: 4
  \item Gaussian width: 0.3 \AA
  \end{itemize}
\item \textbf{Mechanical response}: Numerical force derivatives with respect to 0.01 \AA{} atomic displacements
\end{enumerate}

These features were concatenated into a single vector per configuration, normalized, and projected using t-SNE (perplexity=30, learning rate=200) for visualization.

\subsection{Evaluation Metrics Definitions}
\begin{itemize}
\item \textbf{Force MAE}: $\frac{1}{3N} \sum_{i=1}^{N} \sum_{\alpha=x,y,z} |F_{i,\alpha}^{\text{MLFF}} - F_{i,\alpha}^{\text{DFT}}|$
\item \textbf{Energy MAE}: $\frac{1}{N} |E^{\text{MLFF}} - E^{\text{DFT}}|$
\item \textbf{RMSD Growth Rate}: Linear fit slope of $\text{RMSD}(t) = \sqrt{\frac{1}{N}\sum_{i=1}^{N}|\mathbf{r}_i(t) - \mathbf{r}_i(0)|^2}$
\item \textbf{Silhouette Score}: Average of $\frac{b_i - a_i}{\max(a_i, b_i)}$ where $a_i$ is mean intra-cluster distance and $b_i$ is mean nearest-cluster distance
\end{itemize}

\subsection{Detailed Migration Pathway Analysis}
\subsubsection{In-Gap Diffusion}
This pathway tests interpolative accuracy, as the atomic environments remain similar to equilibrium configurations in the training data. The clear performance hierarchy---with fine-tuned models excelling and scratch failing---validates that fine-tuning successfully ``sharpens'' the foundation model's PES for in-domain predictions.

\subsubsection{Deep Penetration}
This pathway creates severe lattice distortions as Cr pushes through covalently bonded layers, accessing configurations far from the training distribution. The universal failure at transition states, despite accurate endpoint predictions, reveals how models trained on near-equilibrium data develop strong inductive biases that break down for strained configurations.

The scratch model's accidentally lower error on this OOD task is particularly instructive. Its globally inaccurate PES happens to be less catastrophically wrong in this specific high-energy region---not through physical insight but random chance. This underscores that the foundation model's ``general'' knowledge contains implicit assumptions about equilibrium physics that fail dramatically under extrapolation.
\subsubsection{Implications for MLFF Development}

These findings highlight critical considerations for practical MLFF deployment:
\begin{itemize}
    \item \textbf{Validation Insufficiency}: Testing only on stable configurations masks critical failures at transition states
    \item \textbf{Hidden Extrapolation}: Models can appear accurate at trajectory endpoints while failing at crucial transition states
    \item \textbf{Data Quality over Quantity}: Task-specific training data outweighs larger but less relevant datasets
    \item \textbf{Foundation Model Limitations}: Pre-trained models require careful adaptation for processes involving significant atomic rearrangement
\end{itemize}

The results motivate advanced strategies like uncertainty-guided active learning that can identify high-uncertainty regions and intelligently augment training sets for improved extrapolative performance.

\subsection{Nudged Elastic Band (NEB) Benchmark for Cr Migration}

\begin{figure}[htbp]
    \centering
    \includegraphics[width=0.8\textwidth]{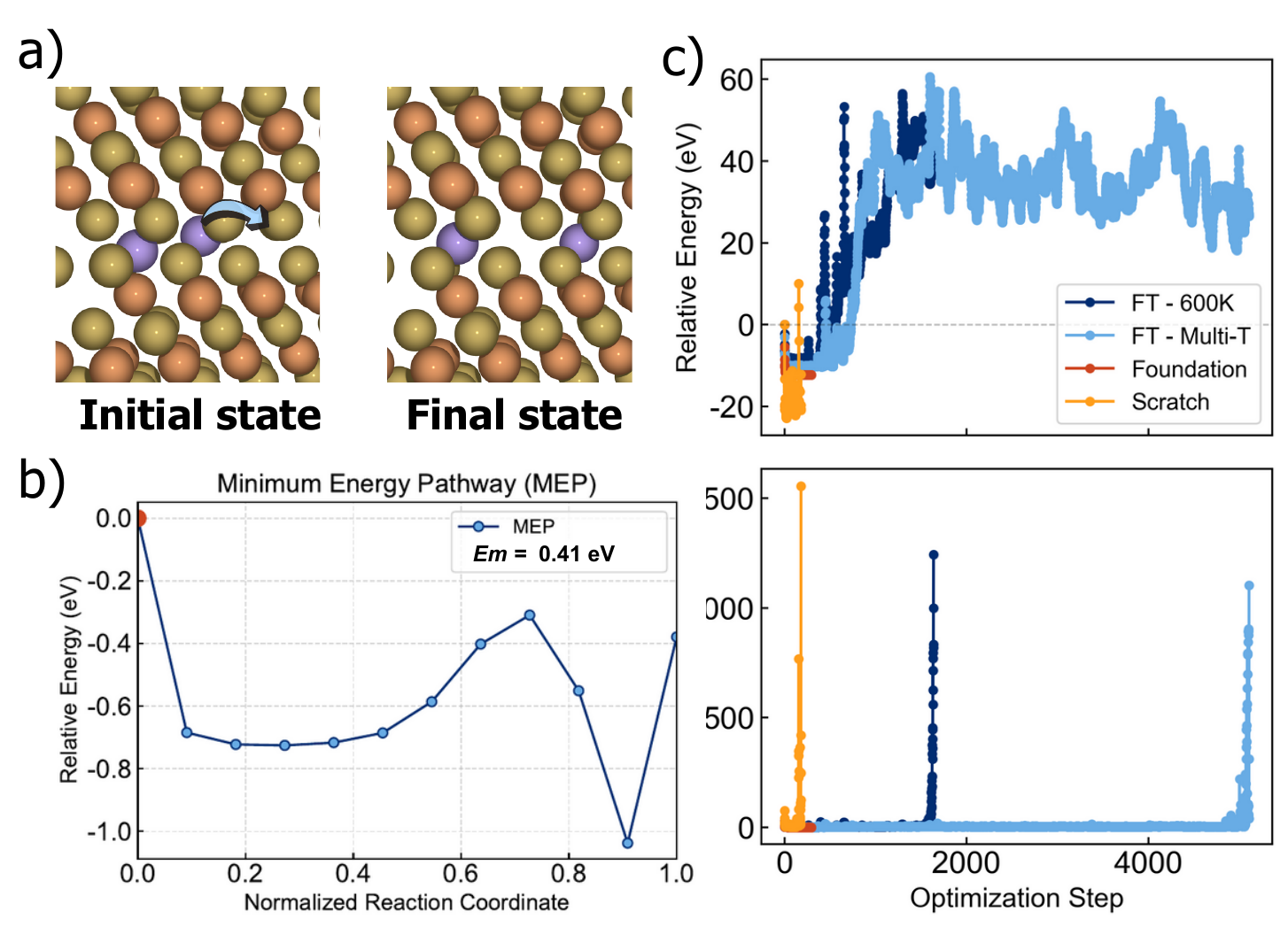}
    \caption{Nudged Elastic Band (NEB) benchmark of MACE models for a Cr atom migration. (a) Visualization of the initial and final states of the diffusion pathway. (b) The converged Minimum Energy Pathway (MEP) calculated with the MACE foundation model, yielding a migration barrier of 0.41 eV. (c) The evolution of the relative system energy (top panel) and the maximum force ($f_{max}$, bottom panel) during the NEB optimization for each model. The foundation model (red) converges smoothly. In contrast, the scratch (orange), FT-600K (dark blue), and FT-MultiT (light blue) models all exhibit explosive behavior, where a rapid increase in $f_{max}$ indicates instability and leads to the termination of the calculation.}
    \label{fig:neb_benchmark}
\end{figure}

To directly assess the practical performance of each model on a physically critical task, we performed a Nudged Elastic Band (NEB) calculation for a Cr atom migration event, as shown in Figure~\ref{fig:neb_benchmark}. This calculation serves as an effective and challenging probe to distinguish the models' stability and predictive accuracy when exploring unseen transition-state geometries. The results reveal a dramatic difference in performance that is not apparent from the RMSE metrics alone.

Both the model trained from scratch and the fine-tuned models (FT-600K and FT-MultiT) exhibit explosive behavior during the NEB optimization, as evidenced by the sudden spike in the maximum force ($f_{max}$) shown in Figure~\ref{fig:neb_benchmark}c. This instability forced the early termination of the calculations. This failure suggests that the interpolated images along the NEB path introduced metastable configurations, such as an unphysical separation of the material layers, which were outside the manifold of the training data. For the fine-tuned models, this indicates a form of catastrophic forgetting, where the models lost their generalized stability after being trained on a narrow dataset.

In contrast, the MACE foundation model, without any system-specific fine-tuning, successfully converged the NEB calculation. It predicts a migration barrier of 0.41 eV (Figure~\ref{fig:neb_benchmark}b), a value in good agreement with DFT calculations for similar in-gap diffusion pathways (~0.3 eV). This level of accuracy, close to the bounds of chemical accuracy, demonstrates the foundation model's exceptional capability to generalize to complex transition-state configurations. This benchmark underscores that evaluating performance on dynamic, physically relevant processes is a critical and necessary step for validating the true capabilities of MLFFs.

\bibliography{neurips_2025}
\bibliographystyle{unsrtnat}






\newpage
\section*{NeurIPS Paper Checklist}
\begin{ack}

The checklist is designed to encourage best practices for responsible machine learning research, addressing issues of reproducibility, transparency, research ethics, and societal impact. Do not remove the checklist: {\bf The papers not including the checklist will be desk rejected.} The checklist should follow the references and follow the (optional) supplemental material.  The checklist does NOT count towards the page
limit. 

Please read the checklist guidelines carefully for information on how to answer these questions. For each question in the checklist:
\begin{itemize}
    \item You should answer \answerYes{}, \answerNo{}, or \answerNA{}.
    \item \answerNA{} means either that the question is Not Applicable for that particular paper or the relevant information is Not Available.
    \item Please provide a short (1–2 sentence) justification right after your answer (even for NA). 
\end{itemize}

{\bf The checklist answers are an integral part of your paper submission.} They are visible to the reviewers, area chairs, senior area chairs, and ethics reviewers. You will be asked to also include it (after eventual revisions) with the final version of your paper, and its final version will be published with the paper.

The reviewers of your paper will be asked to use the checklist as one of the factors in their evaluation. While "\answerYes{}" is generally preferable to "\answerNo{}", it is perfectly acceptable to answer "\answerNo{}" provided a proper justification is given (e.g., "error bars are not reported because it would be too computationally expensive" or "we were unable to find the license for the dataset we used"). In general, answering "\answerNo{}" or "\answerNA{}" is not grounds for rejection. While the questions are phrased in a binary way, we acknowledge that the true answer is often more nuanced, so please just use your best judgment and write a justification to elaborate. All supporting evidence can appear either in the main paper or the supplemental material, provided in appendix. If you answer \answerYes{} to a question, in the justification please point to the section(s) where related material for the question can be found.

IMPORTANT, please:
\begin{itemize}
    \item {\bf Delete this instruction block, but keep the section heading ``NeurIPS Paper Checklist"},
    \item  {\bf Keep the checklist subsection headings, questions/answers and guidelines below.}
    \item {\bf Do not modify the questions and only use the provided macros for your answers}.
\end{itemize} 
 \end{ack}


\begin{itemize}

\item {\bf Claims}
    \item[] Question: Do the main claims made in the abstract and introduction accurately reflect the paper's contributions and scope?
\item[] Answer: \answerYes{}
\item[] Justification: The abstract and introduction clearly state our claims about benchmarking specialist vs. generalist MLFF training strategies for Cr-doped Sb$_2$Te$_3$, and these are fully supported by our systematic evaluation of five distinct training approaches across equilibrium and non-equilibrium properties.
    \item[] Guidelines:
    \begin{itemize}
        \item The answer NA means that the abstract and introduction do not include the claims made in the paper.
        \item The abstract and/or introduction should clearly state the claims made, including the contributions made in the paper and important assumptions and limitations. A No or NA answer to this question will not be perceived well by the reviewers. 
        \item The claims made should match theoretical and experimental results, and reflect how much the results can be expected to generalize to other settings. 
        \item It is fine to include aspirational goals as motivation as long as it is clear that these goals are not attained by the paper. 
    \end{itemize}

\item {\bf Limitations}
    \item[] Question: Does the paper discuss the limitations of the work performed by the authors?
\item[] Answer: \answerYes{}
\item[] Justification: We explicitly discuss limitations throughout the paper, including the single test case nature of our study (Cr-Sb$_2$Te$_3$), the computational constraints that limited MD trajectory lengths, and the fundamental limitations of local-descriptor-based MLFFs for capturing long-range physics (Section 4.5).
    \item[] Guidelines:
    \begin{itemize}
        \item The answer NA means that the paper has no limitation while the answer No means that the paper has limitations, but those are not discussed in the paper. 
        \item The authors are encouraged to create a separate "Limitations" section in their paper.
        \item The paper should point out any strong assumptions and how robust the results are to violations of these assumptions (e.g., independence assumptions, noiseless settings, model well-specification, asymptotic approximations only holding locally). The authors should reflect on how these assumptions might be violated in practice and what the implications would be.
        \item The authors should reflect on the scope of the claims made, e.g., if the approach was only tested on a few datasets or with a few runs. In general, empirical results often depend on implicit assumptions, which should be articulated.
        \item The authors should reflect on the factors that influence the performance of the approach. For example, a facial recognition algorithm may perform poorly when image resolution is low or images are taken in low lighting. Or a speech-to-text system might not be used reliably to provide closed captions for online lectures because it fails to handle technical jargon.
        \item The authors should discuss the computational efficiency of the proposed algorithms and how they scale with dataset size.
        \item If applicable, the authors should discuss possible limitations of their approach to address problems of privacy and fairness.
        \item While the authors might fear that complete honesty about limitations might be used by reviewers as grounds for rejection, a worse outcome might be that reviewers discover limitations that aren't acknowledged in the paper. The authors should use their best judgment and recognize that individual actions in favor of transparency play an important role in developing norms that preserve the integrity of the community. Reviewers will be specifically instructed to not penalize honesty concerning limitations.
    \end{itemize}

\item {\bf Theory assumptions and proofs}
    \item[] Question: For each theoretical result, does the paper provide the full set of assumptions and a complete (and correct) proof?
\item[] Answer: \answerNA{}
\item[] Justification: This paper is an empirical benchmarking study of machine learning force fields and does not contain theoretical results or proofs.
    \item[] Guidelines:
    \begin{itemize}
        \item The answer NA means that the paper does not include theoretical results. 
        \item All the theorems, formulas, and proofs in the paper should be numbered and cross-referenced.
        \item All assumptions should be clearly stated or referenced in the statement of any theorems.
        \item The proofs can either appear in the main paper or the supplemental material, but if they appear in the supplemental material, the authors are encouraged to provide a short proof sketch to provide intuition. 
        \item Inversely, any informal proof provided in the core of the paper should be complemented by formal proofs provided in appendix or supplemental material.
        \item Theorems and Lemmas that the proof relies upon should be properly referenced. 
    \end{itemize}

    \item {\bf Experimental result reproducibility}
    \item[] Question: Does the paper fully disclose all the information needed to reproduce the main experimental results of the paper to the extent that it affects the main claims and/or conclusions of the paper (regardless of whether the code and data are provided or not)?
\item[] Answer: \answerYes{}
\item[] Justification: We provide comprehensive details including DFT parameters (Section 3.1), MACE architecture specifications, training hyperparameters (Section 3.2), MD simulation protocols (Section 3.3), and detailed descriptions of all evaluation metrics and analysis methods.
    \item[] Guidelines:
    \begin{itemize}
        \item The answer NA means that the paper does not include experiments.
        \item If the paper includes experiments, a No answer to this question will not be perceived well by the reviewers: Making the paper reproducible is important, regardless of whether the code and data are provided or not.
        \item If the contribution is a dataset and/or model, the authors should describe the steps taken to make their results reproducible or verifiable. 
        \item Depending on the contribution, reproducibility can be accomplished in various ways. For example, if the contribution is a novel architecture, describing the architecture fully might suffice, or if the contribution is a specific model and empirical evaluation, it may be necessary to either make it possible for others to replicate the model with the same dataset, or provide access to the model. In general. releasing code and data is often one good way to accomplish this, but reproducibility can also be provided via detailed instructions for how to replicate the results, access to a hosted model (e.g., in the case of a large language model), releasing of a model checkpoint, or other means that are appropriate to the research performed.
        \item While NeurIPS does not require releasing code, the conference does require all submissions to provide some reasonable avenue for reproducibility, which may depend on the nature of the contribution. For example
        \begin{enumerate}
            \item If the contribution is primarily a new algorithm, the paper should make it clear how to reproduce that algorithm.
            \item If the contribution is primarily a new model architecture, the paper should describe the architecture clearly and fully.
            \item If the contribution is a new model (e.g., a large language model), then there should either be a way to access this model for reproducing the results or a way to reproduce the model (e.g., with an open-source dataset or instructions for how to construct the dataset).
            \item We recognize that reproducibility may be tricky in some cases, in which case authors are welcome to describe the particular way they provide for reproducibility. In the case of closed-source models, it may be that access to the model is limited in some way (e.g., to registered users), but it should be possible for other researchers to have some path to reproducing or verifying the results.
        \end{enumerate}
    \end{itemize}

\item {\bf Open access to data and code}
    \item[] Question: Does the paper provide open access to the data and code, with sufficient instructions to faithfully reproduce the main experimental results, as described in supplemental material?
\item[] Answer: \answerYes{}
\item[] Justification: We provide detailed methodology for reproducibility. The code and datasets are released on our Github repository (\hyperlink{https://github.com/yicao-elina/mlff-benchmark-cr-sb2te3.git}{mlff-benchmark-cr-sb2te3}), which is actively maintained and will be updated with the newest code and results. Additionally, we use publicly available tools (MACE, Quantum Espresso, ASE) and provide sufficient detail in both the main text and appendix for independent reproduction.
    \item[] Guidelines:
    \begin{itemize}
        \item The answer NA means that paper does not include experiments requiring code.
        \item Please see the NeurIPS code and data submission guidelines (\url{https://nips.cc/public/guides/CodeSubmissionPolicy}) for more details.
        \item While we encourage the release of code and data, we understand that this might not be possible, so “No” is an acceptable answer. Papers cannot be rejected simply for not including code, unless this is central to the contribution (e.g., for a new open-source benchmark).
        \item The instructions should contain the exact command and environment needed to run to reproduce the results. See the NeurIPS code and data submission guidelines (\url{https://nips.cc/public/guides/CodeSubmissionPolicy}) for more details.
        \item The authors should provide instructions on data access and preparation, including how to access the raw data, preprocessed data, intermediate data, and generated data, etc.
        \item The authors should provide scripts to reproduce all experimental results for the new proposed method and baselines. If only a subset of experiments are reproducible, they should state which ones are omitted from the script and why.
        \item At submission time, to preserve anonymity, the authors should release anonymized versions (if applicable).
        \item Providing as much information as possible in supplemental material (appended to the paper) is recommended, but including URLs to data and code is permitted.
    \end{itemize}

\item {\bf Experimental setting/details}
    \item[] Question: Does the paper specify all the training and test details (e.g., data splits, hyperparameters, how they were chosen, type of optimizer, etc.) necessary to understand the results?
\item[] Answer: \answerYes{}
\item[] Justification: Section 3.2 and the Appendix provide complete training details including optimizer (Adam), learning rate ($1\times10^{-3}$), batch size (4), early stopping criteria, temperature settings for AIMD (300K, 600K, 1200K), and dataset sizes (~20,000 configurations).
    \item[] Guidelines:
    \begin{itemize}
        \item The answer NA means that the paper does not include experiments.
        \item The experimental setting should be presented in the core of the paper to a level of detail that is necessary to appreciate the results and make sense of them.
        \item The full details can be provided either with the code, in appendix, or as supplemental material.
    \end{itemize}

\item {\bf Experiment statistical significance}
    \item[] Question: Does the paper report error bars suitably and correctly defined or other appropriate information about the statistical significance of the experiments?
    \item[] Answer: \answerYes{}
    \item[] Justification: 
    For model training, I used three different random seeds to verify the consistency of model behavior. The performance was not affected by the seeds, with all models showing nearly identical results. 
    
    For the NEB diffusion pathways, we conducted multiple independent calculations with different initial perturbations of the migration pathway to establish statistical significance. These results are presented with proper error bars in Figure \ref{fig:barrier}b, quantifying the uncertainty in migration energy predictions across all MLFF training strategies. 
    
    For thermodynamic and structural properties, we analyzed extended trajectories (200 ps) to ensure proper equilibration and sampling. While computational constraints limited us to single training runs for each MLFF strategy, we verified consistency by computing time-averaged properties.
    \item[] Guidelines:
    \begin{itemize}
        \item The answer NA means that the paper does not include experiments.
        \item The authors should answer "Yes" if the results are accompanied by error bars, confidence intervals, or statistical significance tests, at least for the experiments that support the main claims of the paper.
        \item The factors of variability that the error bars are capturing should be clearly stated (for example, train/test split, initialization, random drawing of some parameter, or overall run with given experimental conditions).
        \item The method for calculating the error bars should be explained (closed form formula, call to a library function, bootstrap, etc.)
        \item The assumptions made should be given (e.g., Normally distributed errors).
        \item It should be clear whether the error bar is the standard deviation or the standard error of the mean.
        \item It is OK to report 1-sigma error bars, but one should state it. The authors should preferably report a 2-sigma error bar than state that they have a 96\% CI, if the hypothesis of Normality of errors is not verified.
        \item For asymmetric distributions, the authors should be careful not to show in tables or figures symmetric error bars that would yield results that are out of range (e.g. negative error rates).
        \item If error bars are reported in tables or plots, The authors should explain in the text how they were calculated and reference the corresponding figures or tables in the text.
    \end{itemize}

\item {\bf Experiments compute resources}
    \item[] Question: For each experiment, does the paper provide sufficient information on the computer resources (type of compute workers, memory, time of execution) needed to reproduce the experiments?
    \item[] Answer: \answerYes{} 
    \item[] Justification: The paper provides detailed information on the compute resources used for all experiments, including hardware type (GPU), execution time, system sizes, and number of trajectories/configurations. The compute requirements are summarized in Table~\ref{tab:compute}. This includes AIMD, nudged elastic band (NEB), and machine-learning force field (MLFF) MD runs. Both per-trajectory costs and total project costs are explicitly stated, allowing independent researchers to reproduce the experiments and estimate compute requirements. Furthermore, the sbatch script for the MLFF-MD run explicitly discloses the resource request (1 A100 GPU, 4 MPI tasks, 8 CPUs per task), ensuring clarity in execution environment specification. The disclosure includes total wall-time estimates across configurations and temperatures, thereby providing a transparent account of the actual compute footprint of the research project.

\begin{table}[htbp]
    \centering
    \small
    \caption{Summary of compute resources required for different experiments.}
    \label{tab:compute}
    \begin{tabular}{lccc}
        \toprule
        Experiment & System / Tasks & Hardware & Compute Cost \\
        \midrule
        AIMD & 120 atoms; 10 configs $\times$ 3 temps & 1 GPU (A100) & 9 days / traj (270 runs; $\sim$2430 GPU-days) \\
        NEB & 60 atoms (8 traj); 120 atoms (2 traj) & 1 GPU (A100) & 6 days / traj (10 runs; $\sim$60 GPU-days) \\
        MLFF-MD & 2050 atoms; 4 models $\times$ 2 tasks & 1 GPU (A100) & $\sim$2 hrs / run (8 runs; $\sim$16 GPU-hours) \\
        \bottomrule
    \end{tabular}
\end{table}

    \item[] Guidelines:
    \begin{itemize}
        \item The answer NA means that the paper does not include experiments.
        \item The paper should indicate the type of compute workers CPU or GPU, internal cluster, or cloud provider, including relevant memory and storage.
        \item The paper should provide the amount of compute required for each of the individual experimental runs as well as estimate the total compute. 
        \item The paper should disclose whether the full research project required more compute than the experiments reported in the paper (e.g., preliminary or failed experiments that didn't make it into the paper). 
    \end{itemize}
    
\item {\bf Code of ethics}
    \item[] Question: Does the research conducted in the paper conform, in every respect, with the NeurIPS Code of Ethics \url{https://neurips.cc/public/EthicsGuidelines}?
\item[] Answer: \answerYes{}
\item[] Justification: Our research on machine learning force fields for materials science applications adheres to all ethical guidelines, involves no human subjects, and presents honest assessments of both capabilities and limitations.
    \item[] Guidelines:
    \begin{itemize}
        \item The answer NA means that the authors have not reviewed the NeurIPS Code of Ethics.
        \item If the authors answer No, they should explain the special circumstances that require a deviation from the Code of Ethics.
        \item The authors should make sure to preserve anonymity (e.g., if there is a special consideration due to laws or regulations in their jurisdiction).
    \end{itemize}

\item {\bf Broader impacts}
    \item[] Question: Does the paper discuss both potential positive societal impacts and negative societal impacts of the work performed?
\item[] Answer: \answerYes{}
\item[] Justification: The paper explicitly addresses both positive and potential negative societal impacts. On the positive side, our work contributes to accelerating materials discovery by providing a benchmark framework that combines MD simulations under operational temperatures with nudged elastic band migration studies. This integrated approach enables simultaneous access to thermodynamic and kinetic insights, while offering a rigorous platform to test interpolation and extrapolation capabilities of machine-learned force fields (MLFFs). Importantly, the benchmark pipeline is not limited to the 2D materials investigated here: it can be extended to a wide range of systems with different dopants or migrating species. Such generalizability allows for faster and more reliable evaluation of MLFFs, thereby shortening the model improvement cycle and supporting the development of more robust and flexibly tunable fine-tuning strategies. Ultimately, this work has the potential to significantly boost computational efficiency and accelerate the pace of scientific discovery in materials science.

On the negative side, while direct societal risks are minimal given the fundamental nature of the research, there are indirect considerations. More accurate and efficient MLFFs could accelerate the discovery of materials with dual-use potential, including those relevant for defense or energy storage in sensitive contexts. Acknowledging this possibility underscores the importance of ensuring that such computational advancements are guided by responsible dissemination and ethical application.

    \item[] Guidelines:
    \begin{itemize}
        \item The answer NA means that there is no societal impact of the work performed.
        \item If the authors answer NA or No, they should explain why their work has no societal impact or why the paper does not address societal impact.
        \item Examples of negative societal impacts include potential malicious or unintended uses (e.g., disinformation, generating fake profiles, surveillance), fairness considerations (e.g., deployment of technologies that could make decisions that unfairly impact specific groups), privacy considerations, and security considerations.
        \item The conference expects that many papers will be foundational research and not tied to particular applications, let alone deployments. However, if there is a direct path to any negative applications, the authors should point it out. For example, it is legitimate to point out that an improvement in the quality of generative models could be used to generate deepfakes for disinformation. On the other hand, it is not needed to point out that a generic algorithm for optimizing neural networks could enable people to train models that generate Deepfakes faster.
        \item The authors should consider possible harms that could arise when the technology is being used as intended and functioning correctly, harms that could arise when the technology is being used as intended but gives incorrect results, and harms following from (intentional or unintentional) misuse of the technology.
        \item If there are negative societal impacts, the authors could also discuss possible mitigation strategies (e.g., gated release of models, providing defenses in addition to attacks, mechanisms for monitoring misuse, mechanisms to monitor how a system learns from feedback over time, improving the efficiency and accessibility of ML).
    \end{itemize}
    
\item {\bf Safeguards}
    \item[] Question: Does the paper describe safeguards that have been put in place for responsible release of data or models that have a high risk for misuse (e.g., pretrained language models, image generators, or scraped datasets)?
\item[] Answer: \answerNA{}
\item[] Justification: Our work on force fields for materials simulation poses no direct risks for misuse. The models are specific to Cr-Sb$_2$Te$_3$ systems and have no applications outside scientific research.
    \item[] Guidelines:
    \begin{itemize}
        \item The answer NA means that the paper poses no such risks.
        \item Released models that have a high risk for misuse or dual-use should be released with necessary safeguards to allow for controlled use of the model, for example by requiring that users adhere to usage guidelines or restrictions to access the model or implementing safety filters. 
        \item Datasets that have been scraped from the Internet could pose safety risks. The authors should describe how they avoided releasing unsafe images.
        \item We recognize that providing effective safeguards is challenging, and many papers do not require this, but we encourage authors to take this into account and make a best faith effort.
    \end{itemize}

\item {\bf Licenses for existing assets}
    \item[] Question: Are the creators or original owners of assets (e.g., code, data, models), used in the paper, properly credited and are the license and terms of use explicitly mentioned and properly respected?
\item[] Answer: \answerYes{}
\item[] Justification: We properly cite MACE (ref 12), MACE-MP foundation model, Quantum Espresso, and ASE. We use these tools according to their open-source licenses, though specific license details could be made more explicit.
    \item[] Guidelines:
    \begin{itemize}
        \item The answer NA means that the paper does not use existing assets.
        \item The authors should cite the original paper that produced the code package or dataset.
        \item The authors should state which version of the asset is used and, if possible, include a URL.
        \item The name of the license (e.g., CC-BY 4.0) should be included for each asset.
        \item For scraped data from a particular source (e.g., website), the copyright and terms of service of that source should be provided.
        \item If assets are released, the license, copyright information, and terms of use in the package should be provided. For popular datasets, \url{paperswithcode.com/datasets} has curated licenses for some datasets. Their licensing guide can help determine the license of a dataset.
        \item For existing datasets that are re-packaged, both the original license and the license of the derived asset (if it has changed) should be provided.
        \item If this information is not available online, the authors are encouraged to reach out to the asset's creators.
    \end{itemize}

\item {\bf New assets}
    \item[] Question: Are new assets introduced in the paper well documented and is the documentation provided alongside the assets?
\item[] Answer: \answerNA{}
\item[] Justification: Although we train new models as part of the study, the primary contribution of the paper lies in the benchmarking methodology and the systematic evaluation of training strategies. The intent is not to introduce a new publicly released asset, but rather to provide a reproducible and extensible evaluation framework. Thus, no new standalone assets are released, and the answer is not applicable in the context of this work.

    \item[] Guidelines:
    \begin{itemize}
        \item The answer NA means that the paper does not release new assets.
        \item Researchers should communicate the details of the dataset/code/model as part of their submissions via structured templates. This includes details about training, license, limitations, etc. 
        \item The paper should discuss whether and how consent was obtained from people whose asset is used.
        \item At submission time, remember to anonymize your assets (if applicable). You can either create an anonymized URL or include an anonymized zip file.
    \end{itemize}

\item {\bf Crowdsourcing and research with human subjects}
    \item[] Question: For crowdsourcing experiments and research with human subjects, does the paper include the full text of instructions given to participants and screenshots, if applicable, as well as details about compensation (if any)? 
\item[] Answer: \answerNA{}
\item[] Justification: This paper involves only computational materials science research with no human subjects or crowdsourcing components.
    \item[] Guidelines:
    \begin{itemize}
        \item The answer NA means that the paper does not involve crowdsourcing nor research with human subjects.
        \item Including this information in the supplemental material is fine, but if the main contribution of the paper involves human subjects, then as much detail as possible should be included in the main paper. 
        \item According to the NeurIPS Code of Ethics, workers involved in data collection, curation, or other labor should be paid at least the minimum wage in the country of the data collector. 
    \end{itemize}

\item {\bf Institutional review board (IRB) approvals or equivalent for research with human subjects}
    \item[] Question: Does the paper describe potential risks incurred by study participants, whether such risks were disclosed to the subjects, and whether Institutional Review Board (IRB) approvals (or an equivalent approval/review based on the requirements of your country or institution) were obtained?
\item[] Answer: \answerNA{}
\item[] Justification: This paper does not involve research with human subjects.
    \item[] Guidelines:
    \begin{itemize}
        \item The answer NA means that the paper does not involve crowdsourcing nor research with human subjects.
        \item Depending on the country in which research is conducted, IRB approval (or equivalent) may be required for any human subjects research. If you obtained IRB approval, you should clearly state this in the paper. 
        \item We recognize that the procedures for this may vary significantly between institutions and locations, and we expect authors to adhere to the NeurIPS Code of Ethics and the guidelines for their institution. 
        \item For initial submissions, do not include any information that would break anonymity (if applicable), such as the institution conducting the review.
    \end{itemize}

\item {\bf Declaration of LLM usage}
    \item[] Question: Does the paper describe the usage of LLMs if it is an important, original, or non-standard component of the core methods in this research? Note that if the LLM is used only for writing, editing, or formatting purposes and does not impact the core methodology, scientific rigorousness, or originality of the research, declaration is not required.
    \item[] Answer: \answerNA{}
    \item[] Justification: No large language models (LLMs) were used as part of the research methodology. The core contributions rely exclusively on physics-based machine learning force fields (MACE architecture) trained on density functional theory (DFT) data. Since LLMs did not play any role in method development, experimentation, benchmarking, or scientific analysis, their declaration is not applicable in the context of this work.

    \item[] Guidelines:
    \begin{itemize}
        \item The answer NA means that the core method development in this research does not involve LLMs as any important, original, or non-standard components.
        \item Please refer to our LLM policy (\url{https://neurips.cc/Conferences/2025/LLM}) for what should or should not be described.
    \end{itemize}

\end{itemize}

\end{document}